\documentclass[aps,prl,reprint,superscriptaddress,nofootinbib]{revtex4-2}
\usepackage[T1]{fontenc}
\usepackage[utf8]{inputenc}
\usepackage{amsmath,amssymb,mathtools,bm,braket}
\usepackage{graphicx}
\usepackage{array,booktabs}
\usepackage{xcolor}
\usepackage{tikz}
\usetikzlibrary{arrows.meta,positioning,calc,fit}
\usepackage[colorlinks=true,citecolor=blue,linkcolor=blue,urlcolor=blue]{hyperref}

\newcommand{\Zd}{\mathbb{Z}_d}
\newcommand{\rank}{\operatorname{rank}}

\newcommand{\Tr}{\operatorname{Tr}}

\begin{document}


\title{Auxiliary Schmidt Rank as a Resource for Photonic Bell Measurements}


\author{Pradip Laha}
\email{plaha@uni-mainz.de}
\affiliation{Institute of Physics, Johannes Gutenberg-Universit\"at Mainz, Staudingerweg 7, 55128 Mainz, Germany}
\author{Peter van Loock}
\email{loock@uni-mainz.de}
\affiliation{Institute of Physics, Johannes Gutenberg-Universit\"at Mainz, Staudingerweg 7, 55128 Mainz, Germany}


\begin{abstract}


In quantum communication and fusion-based quantum computation, photonic Bell measurements are fundamentally limited when only passive linear optics is employed. While for qubits, some Bell states can be unambiguously identified with static beam splitters and no extra photons or entanglement, additional auxiliary photons or at least additional auxiliary degrees of freedom with a certain level of additional entanglement are needed to approach or attain a complete, deterministic Bell measurement. Here, we prove an exact resource threshold when the same two photons carry system qudits of dimension $d$ and a fixed auxiliary entangled state $\Phi$, possibly distributed over several additional degrees of freedom, with total Schmidt rank $r_\Phi$. We show that a single conclusive Bell-label functional can occur for $r_\Phi\geqslant\lceil d/2\rceil$, but deterministic discrimination of all $d^2$ Bell-state labels requires $r_\Phi\geqslant d$. A maximally entangled rank-$d$ auxiliary state achieves the bound by local Bell-basis sorting between each photon's system and auxiliary degrees of freedom. Thus, the auxiliary Schmidt rank is a certified resource for ancilla-photon-free, embedded photonic Bell measurements.   

\end{abstract}


\maketitle

\emph{Introduction.--} Bell-state or simply Bell measurements (BMs) often set the usable success rate of photonic quantum protocols~\cite{BianchiReview2026}. They provide the joint two-photon readout behind teleportation~\cite{Bennett1993,Bouwmeester1997}, entanglement swapping~\cite{Zukowski1993,Pan1998}, dense coding~\cite{Bennett1992,Mattle1996}, quantum repeaters and networks~\cite{Briegel1998,Kimble2008,Wehner2018}, and fusion-based photonic quantum computation~\cite{Raussendorf2001,Knill2001,Browne2005,Kok2007}. In an abstract circuit model, a BM is simply a projective measurement in an entangled basis. In optics, the same statement hides a resource question: how many photons are actually populated, which auxiliary degrees of freedom are occupied, and which operations are allowed before detection?

The canonical two-photon polarization two-qubit Bell-state analyzer illustrates this resource dependence. With two photons, passive linear optics, vacuum ancillary modes, and photon-number-resolving (PNR) detection, one can unambiguously identify only part of the four polarization Bell states~\cite{Weinfurter1994,Braunstein1995,Michler1996}.  The impossibility of deterministic analysis in this resource class, and the optimal $1/2$ success probability for unambiguous two-qubit Bell-state discrimination are well established~\cite{Vaidman1999,Lutkenhaus1999,Calsamiglia2001}. Importantly, this is not a universal limit on photonic BMs. It is a theorem about one specific resource class. Hyperentanglement-assisted analyzers evade the two-qubit $50\%$ limit by using a fixed auxiliary entangled degree of freedom carried by the same photons~\cite{Kwiat1998,Walborn2003,Schuck2006,Barbieri2007,Wei2007}.  Other approaches leave this resource class: ancillary-photon schemes change the populated Fock sector~\cite{Grice2011,Ewert2014,Olivo2018,Bayerbach2023,Hauser2025,Baghdasar2025}, while predetection squeezing and related active operations modify the measurement algebra~\cite{Zaidi2013,Kilmer2019,Bianchi2025,Bianchi2025_PRA}. Protocol-level demonstrations, such as boosted teleportation, then build on these enhanced BMs~\cite{DAurelio2025}.
 
High-dimensional photonic qudits make this distinction sharper. Path and time-frequency encodings are natural resources for multidimensional quantum networks~\cite{Bacco2021}, whereas orbital angular momentum and transverse mode encodings provide large structured light mode spaces~\cite{Mirhosseini2015,Nape2023}. High-dimensional photonic entanglement can increase information carried per photon and modify loss, noise, and circuit-depth tradeoffs for communication and network tasks~\cite{Cozzolino2019,Erhard2020}. A pair of $d$-dimensional qudits has $d^2$ generalized Bell-state labels, but a larger optical mode space does not by itself supply a larger measurement resource. Calsamiglia's linear-optical bound shows that the Schmidt number of each POVM element is limited by the number of initially populated photons~\cite{Calsamiglia2002}; Du\v{s}ek emphasized the corresponding separation between bare two-qudit discrimination and schemes that add auxiliary resources~\cite{Dusek2001}.  In the strict two-photon/vacuum-mode setting, a microscopic two-click vector has Schmidt rank at most two, whereas every generalized $d$-dimensional Bell state has Schmidt rank $d$. Thus, for $d>2$, the bare high-dimensional problem is not merely an inefficient analogue of the qubit $1/2$ limit: no conclusive generalized Bell-label outcome exists with nonzero probability.

Recent high-dimensional and encoded proposals operate precisely at this boundary. Auxiliary entanglement in additional degrees of freedom can restore deterministic Bell-label readout in ideal constructions~\cite{Zhang2019}; what is not settled by such constructions is whether a lower auxiliary Schmidt rank could ever suffice. Auxiliary populated states and efficient analyzer architectures change the available optical resource~\cite{Bharos2025}. Logical BMs and fusion protocols instead optimize encoded readout, graph-building, or fusion operations, rather than full physical Bell-label reporting~\cite{Lee_PRL_2015,Ewert_PRL_2016,Ewert_PRA_2017,Lee_PRA_2019,Schmidt_PRA_2019,Hilaire2023,Bell_PRXQuantum_2023,Schmidt2024Fusion,Ustun2025,Yamazaki2025,Reiss2026LogicalBM,Laha_HDBM_Sqz_2026}. Predetection squeezing lies outside the passive analyzer model considered here~\cite{Bianchi2025}. Grouped four-dimensional analyzers provide useful partial information without implementing a complete $d^2$-label BM~\cite{Zeng2025}. These developments make the central issue operational: before asking whether a high-dimensional BM is possible, one must say which photons are populated, which auxiliary state is present, which operations are allowed, and whether the goal is complete Bell-label decoding, grouped information, or fusion.

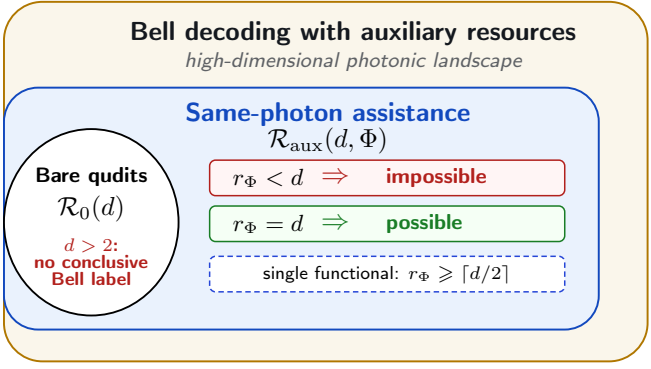
\begin{figure}[t]
\centering
\resizebox{\columnwidth}{!}{%
\begin{tikzpicture}[x=1cm,y=1cm,font=\sffamily]
\definecolor{outergold}{RGB}{176,119,0}
\definecolor{outerfill}{RGB}{250,246,235}
\definecolor{assistblue}{RGB}{18,73,190}
\definecolor{assistfill}{RGB}{234,242,253}
\definecolor{redbox}{RGB}{178,32,28}
\definecolor{greenbox}{RGB}{25,125,38}
\definecolor{dashblue}{RGB}{38,70,215}
\definecolor{ink}{RGB}{20,24,30}
\def\W{8.70}\def\H{4.86}
\draw[rounded corners=0.52cm,line width=0.75pt,draw=outergold,fill=outerfill]
  (0,0) rectangle (\W,\H);
\node[font=\sffamily\bfseries\fontsize{10.2}{10.8}\selectfont,text=ink] at (4.72,4.43)
  {Bell decoding with auxiliary resources};
\node[font=\sffamily\itshape\fontsize{8.0}{8.5}\selectfont,text=ink!68] at (4.72,4.05)
  {high-dimensional photonic landscape};
\draw[rounded corners=0.46cm,line width=0.75pt,draw=assistblue,fill=assistfill]
  (0,0.43) rectangle (8.04,3.70);
\node[text=assistblue,font=\sffamily\bfseries\fontsize{10.2}{10.8}\selectfont] at (4.38,3.34)
  {Same-photon assistance};
\node[font=\sffamily\fontsize{10.0}{10.5}\selectfont] at (4.38,2.98)
  {$\mathcal{R}_{\rm aux}(d,\Phi)$};
\draw[line width=0.72pt,draw=black,fill=white]
  (1.18,1.88) ellipse (1.18 and 1.26);
\node[font=\sffamily\bfseries\fontsize{8.1}{8.6}\selectfont,align=center] at (1.18,2.50)
  {Bare qudits};
\node[font=\sffamily\fontsize{9.7}{10.2}\selectfont] at (1.18,2.06)
  {$\mathcal{R}_{0}(d)$};
\node[text=redbox,font=\sffamily\bfseries\fontsize{6.8}{7.2}\selectfont,align=center] at (1.18,1.35)
  {$d>2$:\\[-0.2mm] no conclusive\\[-0.2mm] Bell label};
\tikzset{stmt/.style={rounded corners=0.08cm,line width=0.58pt,minimum height=0.45cm,align=center,inner sep=1.2pt}}
\def\xA{2.78}\def\xB{7.58}
\draw[stmt,draw=redbox,fill=red!4] (\xA,2.24) rectangle (\xB,2.72);
\node[anchor=west,font=\sffamily\fontsize{8.8}{9.3}\selectfont] at (\xA+0.20,2.48) {$r_\Phi<d$};
\node[text=redbox,font=\sffamily\bfseries\fontsize{11.0}{11.5}\selectfont] at (4.48,2.48) {$\Rightarrow$};
\node[text=redbox,font=\sffamily\bfseries\fontsize{8.0}{8.5}\selectfont,anchor=west] at (5.05,2.48) {impossible};
\draw[stmt,draw=greenbox,fill=green!4] (\xA,1.62) rectangle (\xB,2.10);
\node[anchor=west,font=\sffamily\fontsize{8.8}{9.3}\selectfont] at (\xA+0.20,1.86) {$r_\Phi=d$};
\node[text=greenbox,font=\sffamily\bfseries\fontsize{11.0}{11.5}\selectfont] at (4.48,1.86) {$\Rightarrow$};
\node[text=greenbox,font=\sffamily\bfseries\fontsize{8.0}{8.5}\selectfont,anchor=west] at (5.05,1.86) {possible};
\draw[stmt,draw=dashblue,dash pattern=on 2.1pt off 1.35pt,fill=white]
  (\xA,0.94) rectangle (\xB,1.42);
\node[font=\sffamily\fontsize{7.2}{7.7}\selectfont,align=center] at (5.18,1.18)
  {single functional: $r_\Phi\geqslant \lceil d/2\rceil$};
\end{tikzpicture}
}
\caption{Resource hierarchy for deterministic high-dimensional Bell-state decoding. The bare two-photon/vacuum-mode class $\mathcal{R}_{0}(d)$ admits no conclusive generalized Bell-label outcome for $d>2$~\cite{Calsamiglia2002}. Same-photon auxiliary entanglement can realize a single Bell-label functional at $r_{\Phi}\geqslant \lceil d/2\rceil$, but deterministic decoding in the static passive two-photon model requires full auxiliary Schmidt rank.}
\label{fig:taxonomy}
\end{figure}

We therefore consider a passive two-photon node with no additional populated photons and no active or nonlinear operation. The photons that carry the unknown system qudits may also carry a fixed auxiliary entangled state in one or more additional degrees of freedom. Then the operational question is: what total auxiliary Schmidt rank is required to decode every high-dimensional Bell-state label deterministically? The answer is not set by a single successful click pattern. A deterministic photon-counting analyzer must consistently assign all fine-grained outcomes, including events in which both photons exit the same output mode. This additional constraint raises the resource threshold from the single-outcome condition to a deterministic certificate.



We isolate the same-photon-assisted task.  Two photons carry unknown system qudits $S_A, S_B$ and also a fixed auxiliary state $\ket{\Phi}_R$ in one or more additional degrees of freedom, with combined auxiliary spaces $R_A$ and $R_B$.  No additional photons are populated; arbitrary extra modes may be vacuum.  A single passive interferometer is followed by PNR detection and arbitrary classical coarse-graining. Let $r_\Phi$ denote the Schmidt rank of $\ket{\Phi}_R$ across the combined $R_A|R_B$ auxiliary partition; for a product of independent auxiliary entangled states, the ranks multiply. The fine-grained conclusive pattern condition is stated in Supplemental Sec.~\ref{sec:supp_model}. The input ensemble is
\begin{equation}
  \mathcal{S}^{(d)}_{\Phi} = \left\{\ket{\Psi^{(d)}_{pq}}_{S_A S_B}\otimes\ket{\Phi}_{R_A R_B}:p,q\in\Zd\right\},
\end{equation}
and the task is complete deterministic readout of the system Bell-state label $(p,q)$.  Crucially, this is not a complete BM on the enlarged $S\otimes R$ Hilbert space.  The result is the following sharp existential threshold:
\begin{align}
  &r_\Phi<d \Rightarrow \text{complete deterministic decoding is impossible},\nonumber\\[-1mm]
  &r_\Phi=d\quad\text{is achievable for }\ket{\Phi_d}\text{ in ideal modes}.
  \label{eq:mainresult}
\end{align}
The first line is a no-go theorem for every passive interferometer with arbitrary vacuum modes; the second is a saturating construction. 
Figure~\ref{fig:taxonomy} summarizes the nested resource classes and the two rank thresholds.

\emph{Bare two-photon algebra.--} Any pure bare two-qudit state in two input registers $A$ and $B$ can be written as
\begin{equation}
  \ket{C} = \sum_{m,n=0}^{d-1} C_{mn}a_m^\dagger b_n^\dagger\ket{\mathrm{vac}},\qquad \Tr \left(C^\dagger C\right) = 1 .
  \label{eq:ketC}
\end{equation}
We identify \(a_m^\dagger b_n^\dagger\ket{\mathrm{vac}} \equiv \ket{m}_A\ket{n}_B\), where \(m,n\in\mathbb{Z}_d\) label the logical single-photon modes of the two qudits, not photon occupation numbers.
The generalized Bell matrices are
\begin{equation}
  C_{pq} = d^{-1/2}X^p \, Z^q, \qquad  p,q\in\mathbb{Z}_d.
  \label{eq:bellmat}
\end{equation}
Here, the permutation matrix \(X\) and the diagonal matrix \(Z\) act on the single-qudit computational basis as \(X\ket{n}=\ket{n+1}\) and \(Z\ket{n}=\omega^n\ket{n}\), with $\omega=e^{2\pi i/d}$.
Throughout, additions inside computational-basis labels are taken modulo $d$. Together, $p,q\in\Zd$ label the $d^2$ generalized Bell states,
\begin{equation}
  \ket{\Psi_{pq}^{(d)}} = \frac{1}{\sqrt{d}} \sum_{n=0}^{d-1} \omega^{qn}\ket{n+p}_A\ket{n}_B.
  \label{eqn:bell_states_d}
\end{equation} 
Since both $X$ and $Z$ are invertible, every $C_{pq}$ is full rank. Alternatively, every generalized Bell state has Schmidt rank $d$ and is maximally entangled.

Let the populated logical input registers be $A=\{a_0^\dagger,\ldots,a_{d-1}^\dagger\}$ and $B=\{b_0^\dagger,\ldots,b_{d-1}^\dagger\}$. A passive interferometer maps input to output modes as $c_\mu^\dagger=\sum_\ell U_{\mu\ell}a_\ell^\dagger$. 
For a fine-grained PNR outcome, let \((\mu,\nu)\) denote the occupied output modes. We define the associated normalized two-photon output-Fock
bra by
\begin{equation}
{}_{c}\!\bra{\mu,\nu}
\equiv
\frac{\bra{\mathrm{vac}}\,c_\mu c_\nu}
{\sqrt{1+\delta_{\mu\nu}}}.
\label{eq:detectionbra}
\end{equation}
For \(\mu\neq\nu\), this denotes one photon in each of the two output modes; for \(\mu=\nu\), it denotes two photons in output mode \(\mu\).
Define the row restrictions $\alpha_\mu =(U_{\mu,a_0}^*,\ldots,U_{\mu,a_{d-1}}^*)^T$ and $\beta_\mu= (U_{\mu,b_0}^*,\ldots,U_{\mu,b_{d-1}}^*)^T$ which collect the components of the output row $\mu$ on the populated $A$ and $B$ registers. 
On the logical two-qudit subspace, the amplitude of this fine-grained detection event on $\ket{C}$ is
\begin{equation}
   A_{\mu,\nu}(C) = {}_{c}\!\braket{\mu,\nu|C} = \frac{\alpha_\mu^T C\beta_\nu+\alpha_\nu^T C\beta_\mu}{\sqrt{1+\delta_{\mu\nu}}}.
  \label{eq:bareamp}
\end{equation}
Equivalently, Eq.~\eqref{eq:bareamp} can be written as
$ A_{\mu,\nu}(C)=\Tr(P_{\mu\nu}^{T}C)$, where the effective detection matrix is
\begin{equation}
  P_{\mu\nu} = \frac{\alpha_\mu\beta_\nu^T+\alpha_\nu\beta_\mu^T}{\sqrt{1+\delta_{\mu\nu}}}.
  \label{eq:bareP}
\end{equation}
This is a sum of two outer products, and hence \(\rank P_{\mu\nu}\leqslant 2\). Coarse-graining cannot help: if a coarse-grained outcome is conclusive, then every microscopic pattern inside it is zero on non-target labels, and at least one of them has nonzero target probability. Thus, the strict bare model cannot identify a Bell-state label for \(d>2\): each Bell state has Schmidt rank \(d\), whereas every fine-grained two-photon effect has Schmidt rank at most two. Further details are given in Supplemental Sec.~\ref{sec:supp_bare}.

\emph{Single-outcome-assisted threshold.--} Same-photon auxiliary entanglement changes the effective system functional. On the combined auxiliary spaces, write the fixed auxiliary state as
\begin{equation}
  \ket{\Phi}_R = \sum_{a,b}\Phi_{ab}\ket{a}_{R_A}\ket{b}_{R_B},\qquad r_\Phi=\rank\Phi .
\end{equation}
With auxiliary entanglement, the microscopic detector pattern lives on the bigger space $(S_A\otimes R_A)\otimes(S_B\otimes R_B)$. A microscopic two-click vector $P$ on the enlarged space still has Schmidt rank at most two.  The auxiliary state is known. Therefore, each detector pattern induces an effective system-level matrix by contracting over the auxiliary indices,
\begin{align}
  [\Gamma_\Phi(P)]_{ij} &= \sum_{a,b}P_{(i,a),(j,b)}\Phi_{ab}^*,\\
  \langle P|C\otimes\Phi\rangle &= \Tr\left[\Gamma^\dagger_\Phi(P)\, C\right].
\end{align}
$\Gamma_\Phi(P)$ is the effective Bell-label functional seen by the system after the fixed auxiliary state has been inserted.


Since the two-click vector has Schmidt rank at most two, $P = u_1 v_1^T + u_2 v_2^T$. Let the vectors $u_t,v_t$, each of length $d\,r_\Phi$, be reshaped into $d\times r_\Phi$ matrices $U_t, V_t$ on the system and auxiliary indices ($t=1, 2$). Then
\begin{equation}
  \Gamma_\Phi(P) = \sum_{t=1}^2 U_t\, \Phi^*\, V_t^T .
\label{eq:contract}
\end{equation}
Thus, each rank-one term of the enlarged two-click vector can become a system-level matrix of rank up to $r_\Phi$ after contraction with the rank-$r_\Phi$ auxiliary state, resulting in $\rank \Gamma_\Phi(P)\leqslant 2r_\Phi$. Conversely, any $d\times d$ matrix of rank at most $\min(d,2r_\Phi)$ can be factored in this way by splitting its singular values into two groups. Since an outcome conclusive for label $(p,q)$ must have contraction proportional to $C_{pq}$, a single conclusive Bell-label functional exists in this rank relaxation if and only if
\begin{equation}
  r_\Phi\ge\left\lceil \frac d2\right\rceil \,.
\label{eq:singlethreshold}
\end{equation}
This is a real single-outcome assisted threshold, not an artifact: for $d=3$, a rank-two auxiliary state can generate an individual conclusive contraction for each of the nine Bell labels.  What it does not guarantee is that all such contractions can be completed into one deterministic analyzer. The full contraction-space characterization is proved in Supplemental Sec.~\ref{sec:supp_contraction}.

\emph{Deterministic threshold.--} Determinism imposes the missing physical constraint.  In a deterministic photon-counting analyzer, every fine-grained Fock pattern must be assigned unambiguously.  A nonzero microscopic pattern cannot click on two different Bell labels.  Since the matrices $C_{pq}$ form a complete Hilbert-Schmidt basis, any nonzero contraction that vanishes on all non-target labels must be proportional to the corresponding full-rank matrix $C_{pq}$.  
This requirement also applies to same-output-mode events $(\mu,\mu)$, where both photons are detected in one output mode; such events cannot simply be discarded in a deterministic analyzer.

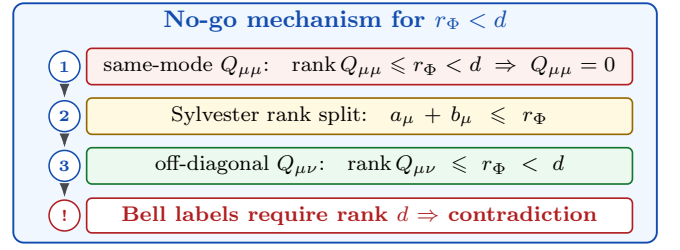
\begin{figure}[t]
\centering
\resizebox{\columnwidth}{!}{%
\begin{tikzpicture}[x=1cm,y=1cm,font=\sffamily,>=Latex]
\definecolor{blue}{RGB}{18,76,175}
\definecolor{softblue}{RGB}{241,247,254}
\definecolor{red}{RGB}{176,30,32}
\definecolor{softred}{RGB}{253,241,240}
\definecolor{gold}{RGB}{150,105,0}
\definecolor{softgold}{RGB}{255,249,225}
\definecolor{green}{RGB}{18,120,45}
\definecolor{softgreen}{RGB}{236,249,238}
\definecolor{ink}{RGB}{24,28,34}

\def\W{8.55}
\def\H{3.18}
\draw[draw=blue,fill=softblue,rounded corners=4pt,line width=0.72pt] (0,0) rectangle (\W,\H);
\node[text=blue,font=\bfseries\fontsize{8.6}{9.1}\selectfont] at (\W/2,2.95)
  {No-go mechanism for $r_\Phi<d$};

\tikzset{
  card/.style={rounded corners=2pt,line width=0.62pt,align=center,inner xsep=5pt,inner ysep=3pt,font=\fontsize{8.1}{8.5}\selectfont},
  flow/.style={-{Latex[length=1.9mm,width=1.35mm]},line width=0.66pt,draw=ink!80,shorten <=0pt,shorten >=0pt},
  circ/.style={circle,minimum size=4.1mm,inner sep=0pt,draw=blue,fill=white,line width=0.58pt,font=\bfseries\fontsize{6.2}{6.2}\selectfont,text=blue}
}

\coordinate (p1) at (0.68,2.34);
\coordinate (p2) at (0.68,1.68);
\coordinate (p3) at (0.68,1.02);
\coordinate (p4) at (0.68,0.36);
\draw[flow] ($(p1)+(0,-0.24)$) -- ($(p2)+(0,0.24)$);
\draw[flow] ($(p2)+(0,-0.24)$) -- ($(p3)+(0,0.24)$);
\draw[flow] ($(p3)+(0,-0.24)$) -- ($(p4)+(0,0.24)$);
\node[circ] at (p1) {1};
\node[circ] at (p2) {2};
\node[circ] at (p3) {3};
\node[circ,draw=red,text=red] at (p4) {!};

\node[card,draw=red,fill=softred,text width=6.90cm] at (4.60,2.34)
 {same-mode $Q_{\mu\mu}$:\quad $\rank Q_{\mu\mu}\leqslant r_\Phi<d\;\Rightarrow\;Q_{\mu\mu}=0$};
\node[card,draw=gold,fill=softgold,text width=6.90cm] at (4.60,1.68)
 {Sylvester rank split:\quad $a_\mu+b_\mu\leqslant r_\Phi$};
\node[card,draw=green,fill=softgreen,text width=6.90cm] at (4.60,1.02)
 {off-diagonal $Q_{\mu\nu}$:\quad $\rank Q_{\mu\nu}\leqslant r_\Phi<d$};
\node[card,draw=red,fill=white,text=red,text width=6.90cm,font=\bfseries\fontsize{8.0}{8.5}\selectfont] at (4.60,0.36)
 {Bell labels require rank $d$ $\Rightarrow$ contradiction};
\end{tikzpicture}
}
\caption{Deterministic no-go mechanism for $r_{\Phi}<d$. Same-mode two-photon outcomes are rank deficient and must vanish in any deterministic analyzer. This enforces the rank split $a_{\mu}+b_{\mu}\leqslant r_{\Phi}$, which makes every off-diagonal contraction rank deficient as well; hence, no fine-grained outcome can carry a Bell-state label.}
\label{fig:proof}
\end{figure}

Put the auxiliary state in Schmidt form on its support,
\begin{equation}
  \ket{\Phi}_R = \sum_{a=1}^{r}\lambda_a\ket{a}_{R_A}\ket{a}_{R_B},\qquad r=r_\Phi,
\end{equation}
with $\Lambda=\operatorname{diag}(\lambda_1,\ldots,\lambda_r)$ invertible.  For each output mode $\mu$, let $X_\mu,Y_\mu\in\mathbb{C}^{d\times r}$ be the projections of the interferometer row onto the populated $A$ and $B$ registers, restricted to the auxiliary Schmidt support.  The contraction for pattern $(\mu,\nu)$ is
\begin{equation}
  Q_{\mu\nu} = \frac{X_\mu\Lambda Y_\nu^T+X_\nu\Lambda Y_\mu^T}{\sqrt{1+\delta_{\mu\nu}}}.
  \label{eq:Qmunu}
\end{equation}
Now assume $r<d$.  For a same-mode event, $Q_{\mu\mu} = \sqrt{2}\,X_\mu\Lambda Y_\mu^T$
has rank at most $r$, so it cannot be proportional to a Bell matrix.  Determinism forces it to vanish: $X_\mu\Lambda Y_\mu^T = 0$ for every $\mu$.
Let $a_\mu=\rank X_\mu$ and $b_\mu=\rank Y_\mu$.  Sylvester's inequality 
gives
\begin{equation}
  0 = \rank(X_\mu\Lambda Y_\mu^T)\geqslant a_\mu+b_\mu-r \,\, \Rightarrow  \, a_\mu+b_\mu\leqslant r .
  \label{eq:ranksplit}
\end{equation}
For an off-diagonal pattern, the rank subadditivity inequality along with Eq.~\eqref{eq:ranksplit} gives
\begin{align}
  \rank Q_{\mu\nu} &\leqslant \rank (X_\mu\Lambda Y_\nu^T) +\rank (X_\nu\Lambda Y_\mu^T)\nonumber\\
  &\leqslant \min(a_\mu,b_\nu)+\min(a_\nu,b_\mu)\nonumber\\
  &\leqslant \frac{a_\mu+b_\nu}{2} + \frac{a_\nu+b_\mu}{2}\leqslant r<d .
  \label{eq:offrank}
\end{align}
Thus, no off-diagonal microscopic contraction can be proportional to a Bell matrix either. If $r_\Phi<d$, no microscopic pattern can be conclusive, contradicting deterministic decoding of all $d^2$ labels. This proves Eq.~\eqref{eq:mainresult}. The proof mechanism is summarized in Fig.~\ref{fig:proof}, and a fully expanded rank argument is given in Supplemental Sec.~\ref{sec:supp_deterministic}. Threshold and finite-resolution detectors are classical mergings of the fine-grained number-resolving patterns. Naturally, the no-go applies to these cases also.

The qutrit boundary case explains the gap between Eqs.~\eqref{eq:mainresult} and \eqref{eq:singlethreshold}.  For $d=3$, auxiliary qubit entanglement has $r_\Phi=2$ and passes the single-outcome test $2r_\Phi\geqslant d$.  Individual conclusive contractions exist; an explicit two-click mode realization for one such qutrit witness is given in Supplemental Sec.~\ref{sec:supp_tightness}.  A deterministic analyzer, however, must also account for every same-mode event.  Those events force the rank split in Eq.~\eqref{eq:ranksplit}, and then Eq.~\eqref{eq:offrank} limits every off-diagonal contraction to rank at most two.  Therefore, auxiliary-qubit assistance is analytically impossible for deterministic qutrit Bell-label decoding in this model.  For $d=4$, both $r_\Phi=2$ and $3$ fail for the same reason; an auxiliary ququart is required.

\emph{Tightness in the ideal mode model.--}
The lower bound is tight in the following existential sense: no auxiliary state with $r_\Phi<d$ can work, while at least one state with $r_\Phi=d$ does. The $d$ Schmidt labels may be physical labels of one auxiliary degree of freedom or composite labels inside several combined auxiliary degrees of freedom. Take
\begin{equation}
\ket{\Phi_d}_{R_A R_B} = \frac{1}{\sqrt{d}}  \sum_{a=0}^{d-1}\ket{a}_{R_A}\ket{a}_{R_B},
\end{equation}
and define, for each photon, the single-photon Bell basis between its system and auxiliary degrees of freedom,
\begin{equation}
\ket{\chi_{mn}}_{SR} = \frac{1}{\sqrt{d}} \sum_{t=0}^{d-1} \omega^{nt}\ket{t}_S\ket{t+m}_R .
\end{equation}
The construction is the ideal-mode realization of the usual entanglement swapping identity, implemented here as local Bell-basis sorting between the system and auxiliary degrees of freedom carried by each photon. With the convention of Eq.~\eqref{eq:bellmat}, one has 
\begin{align}
\ket{\Psi^{(d)}_{pq}}_{S_A S_B}\ket{\Phi_d}_{R_A R_B} &= \frac1d\sum_{m,n\in\Zd}\omega^{-np}
\ket{\chi_{mn}}_{S_A R_A}\nonumber\\
&\qquad\quad\times \ket{\chi_{m+p,q-n}}_{S_B R_B}.
\label{eq:swap}
\end{align}
Therefore local outcomes $(m,n)$ and $(m',n')$ determine
\begin{equation}
p=m'-m, \qquad q=n+n', \pmod d  .
\end{equation}
Since $\{\ket{\chi_{mn}}\}$ is an orthonormal basis of the one-photon $S\otimes R$ mode space, an ideal passive single-particle unitary can map it to distinct output modes \cite{Reck1994,Clements2016}. A block-diagonal interferometer applying this one-particle basis change separately in the $A$ and $B$ spatial registers, followed by PNR detection, decodes all $d^2$ labels deterministically. This is the ideal-mode version of high-dimensional auxiliary-entanglement constructions such as Ref.~\cite{Zhang2019}. Here it serves a different purpose: together with the no-go theorem above, it shows that full auxiliary Schmidt rank is the exact existential threshold. The construction uses no ancillary photons; its nontrivial resource is the full-rank, flat-spectrum auxiliary entangled state. It does not imply that every nonmaximally entangled full-rank auxiliary state is sufficient. The derivation of the swapping identity, the explicit qutrit sorter construction, and the \(d=3,4\) boundary examples can be found in Supplemental Sec.~\ref{sec:supp_tightness}.

For qutrits, this construction is especially transparent.  Label one photon's computational modes by $\ket{s,r}$, $s,r\in\mathbb{Z}_3$.  A local analyzer first sorts by the diagonal index $m=r-s$ and then applies an inverse qutrit Fourier transform on $s$ inside each fixed-$m$ block:
\begin{align}
U_{\rm loc}&=(F_3^\dagger\otimes I_3)\Delta,\\
\langle n,m|U_{\rm loc}|s,r\rangle&=3^{-1/2}\omega^{-ns}\delta_{m,r-s} .
\end{align}
It maps $\ket{\chi_{mn}}$ to the output mode $\ket{n,m}$.  Applying it to both photons gives output pairs $(n_A,m_A)$ and $(n_B,m_B)$ with recovery rule
\begin{equation}
 p=m_B-m_A,  \qquad q=n_A+n_B  \pmod 3.
\end{equation}
Thus, the exact qutrit design rule is $r_\Phi=3$: auxiliary qutrit entanglement suffices in the ideal mode model, while auxiliary-qubit entanglement is impossible. The explicit qutrit sorter construction and its check against the linear-optical distinguishability criteria of Ref.~\cite{vanLoockLutkenhaus2004} are given in Supplemental Sec.~\ref{sec:supp_tightness}.


\emph{Scope and related tasks.--}
The contradiction uses only three ingredients: exactly two populated photons, a fixed auxiliary state carried by those photons, and deterministic assignment of every fine-grained photon-counting pattern to one of the $d^2$ system Bell labels. Empty modes, larger interferometers, and classical postprocessing do not change the rank split because they do not change the two populated photons entering the symmetrized two-click amplitudes.
Schemes with ancillary photons, active or nonlinear operations, logical/fusion measurements, or grouped readout relax one of these assumptions by changing the populated Fock sector, the measurement algebra, or the success criterion. They are, therefore, distinct tasks rather than counterexamples to the present deterministic full-label theorem. Detailed distinctions are summarized in Supplemental Sec.~\ref{sec:supp_scope}.


\emph{Implications.--} 
The theorem makes auxiliary Schmidt rank a certifiable resource. A deterministic high-dimensional decoder using same-photon assistance must specify the total auxiliary Schmidt rank \(r_\Phi\); if \(r_\Phi<d\), the claim is incompatible with static passive optics. Existing deterministic qubit hyperentanglement analyzers are consistent with the threshold because \(r_\Phi=d=2\)~\cite{Schuck2006,Barbieri2007}, while high-dimensional constructions with maximally entangled auxiliary qudits saturate it in the ideal mode model~\cite{Zhang2019}. Protocol-level consequences for teleportation, entanglement swapping, repeaters, dense coding, and fusion-based architectures are summarized in Supplemental Sec.~\ref{sec:supp_protocol}.


Several caveats are part of the result. The sufficiency direction proves the exactness of the Schmidt-rank threshold by example; it does not say that every full-rank {\em nonmaximally} entangled auxiliary state is sufficient. The construction also assumes ideal local control of each photon's $S\otimes R$ one-particle mode space. Decomposing that unitary into experimentally natural elements, with loss, mode mismatch, and detector imperfections included, is a platform-level synthesis problem. Probabilistic and grouped analyzers require separate figures of merit. The present theorem concerns deterministic full-label decoding and explains why the boundary region $\lceil d/2\rceil\leqslant r_\Phi<d$ is deceptive: it contains single conclusive functionals but no complete deterministic analyzer.

In this sense, the result refines the usual no-go language for linear-optical BMs.  It does not say that high-dimensional photonic BMs are impossible.  It says which resource must be present for a specific task.  In the same-photon-assisted two-photon model, total auxiliary Schmidt rank at least \(d\) is not optional: it is necessary for deterministic high-dimensional Bell-label decoding, and a maximally entangled rank-\(d\) auxiliary state shows that the bound is tight in the ideal mode model.

\begin{acknowledgments}
We acknowledge funding from the BMFTR in Germany for support via PhotonQ, QR.N, QuKuK, and QuaPhySI.
\end{acknowledgments}

\bibliography{references}

\begin{thebibliography}{64}%
\makeatletter
\providecommand \@ifxundefined [1]{%
 \@ifx{#1\undefined}
}%
\providecommand \@ifnum [1]{%
 \ifnum #1\expandafter \@firstoftwo
 \else \expandafter \@secondoftwo
 \fi
}%
\providecommand \@ifx [1]{%
 \ifx #1\expandafter \@firstoftwo
 \else \expandafter \@secondoftwo
 \fi
}%
\providecommand \natexlab [1]{#1}%
\providecommand \enquote  [1]{``#1''}%
\providecommand \bibnamefont  [1]{#1}%
\providecommand \bibfnamefont [1]{#1}%
\providecommand \citenamefont [1]{#1}%
\providecommand \href@noop [0]{\@secondoftwo}%
\providecommand \href [0]{\begingroup \@sanitize@url \@href}%
\providecommand \@href[1]{\@@startlink{#1}\@@href}%
\providecommand \@@href[1]{\endgroup#1\@@endlink}%
\providecommand \@sanitize@url [0]{\catcode `\\12\catcode `\$12\catcode
  `\&12\catcode `\#12\catcode `\^12\catcode `\_12\catcode `\%12\relax}%
\providecommand \@@startlink[1]{}%
\providecommand \@@endlink[0]{}%
\providecommand \url  [0]{\begingroup\@sanitize@url \@url }%
\providecommand \@url [1]{\endgroup\@href {#1}{\urlprefix }}%
\providecommand \urlprefix  [0]{URL }%
\providecommand \Eprint [0]{\href }%
\providecommand \doibase [0]{https://doi.org/}%
\providecommand \selectlanguage [0]{\@gobble}%
\providecommand \bibinfo  [0]{\@secondoftwo}%
\providecommand \bibfield  [0]{\@secondoftwo}%
\providecommand \translation [1]{[#1]}%
\providecommand \BibitemOpen [0]{}%
\providecommand \bibitemStop [0]{}%
\providecommand \bibitemNoStop [0]{.\EOS\space}%
\providecommand \EOS [0]{\spacefactor3000\relax}%
\providecommand \BibitemShut  [1]{\csname bibitem#1\endcsname}%
\let\auto@bib@innerbib\@empty
\bibitem [{\citenamefont {Bianchi}\ \emph {et~al.}(2026)\citenamefont
  {Bianchi}, \citenamefont {Marconi},\ and\ \citenamefont
  {Bacco}}]{BianchiReview2026}%
  \BibitemOpen
  \bibfield  {author} {\bibinfo {author} {\bibfnamefont {L.}~\bibnamefont
  {Bianchi}}, \bibinfo {author} {\bibfnamefont {C.}~\bibnamefont {Marconi}},\
  and\ \bibinfo {author} {\bibfnamefont {D.}~\bibnamefont {Bacco}},\ }\bibfield
   {title} {\bibinfo {title} {{B}ell state measurements in quantum optics: a
  review of recent progress and open challenges},\ }\href
  {https://doi.org/10.1088/2058-9565/ae609e} {\bibfield  {journal} {\bibinfo
  {journal} {Quantum Sci. Technol.}\ }\textbf {\bibinfo {volume} {11}},\
  \bibinfo {pages} {023001} (\bibinfo {year} {2026})}\BibitemShut {NoStop}%
\bibitem [{\citenamefont {Bennett}\ \emph {et~al.}(1993)\citenamefont
  {Bennett}, \citenamefont {Brassard}, \citenamefont {Cr\'epeau}, \citenamefont
  {Jozsa}, \citenamefont {Peres},\ and\ \citenamefont
  {Wootters}}]{Bennett1993}%
  \BibitemOpen
  \bibfield  {author} {\bibinfo {author} {\bibfnamefont {C.~H.}\ \bibnamefont
  {Bennett}}, \bibinfo {author} {\bibfnamefont {G.}~\bibnamefont {Brassard}},
  \bibinfo {author} {\bibfnamefont {C.}~\bibnamefont {Cr\'epeau}}, \bibinfo
  {author} {\bibfnamefont {R.}~\bibnamefont {Jozsa}}, \bibinfo {author}
  {\bibfnamefont {A.}~\bibnamefont {Peres}},\ and\ \bibinfo {author}
  {\bibfnamefont {W.~K.}\ \bibnamefont {Wootters}},\ }\bibfield  {title}
  {\bibinfo {title} {Teleporting an unknown quantum state via dual classical
  and {Einstein-Podolsky-Rosen} channels},\ }\href
  {https://doi.org/10.1103/PhysRevLett.70.1895} {\bibfield  {journal} {\bibinfo
   {journal} {Phys. Rev. Lett.}\ }\textbf {\bibinfo {volume} {70}},\ \bibinfo
  {pages} {1895} (\bibinfo {year} {1993})}\BibitemShut {NoStop}%
\bibitem [{\citenamefont {Bouwmeester}\ \emph {et~al.}(1997)\citenamefont
  {Bouwmeester}, \citenamefont {Pan}, \citenamefont {Mattle}, \citenamefont
  {Eibl}, \citenamefont {Weinfurter},\ and\ \citenamefont
  {Zeilinger}}]{Bouwmeester1997}%
  \BibitemOpen
  \bibfield  {author} {\bibinfo {author} {\bibfnamefont {D.}~\bibnamefont
  {Bouwmeester}}, \bibinfo {author} {\bibfnamefont {J.-W.}\ \bibnamefont
  {Pan}}, \bibinfo {author} {\bibfnamefont {K.}~\bibnamefont {Mattle}},
  \bibinfo {author} {\bibfnamefont {M.}~\bibnamefont {Eibl}}, \bibinfo {author}
  {\bibfnamefont {H.}~\bibnamefont {Weinfurter}},\ and\ \bibinfo {author}
  {\bibfnamefont {A.}~\bibnamefont {Zeilinger}},\ }\bibfield  {title} {\bibinfo
  {title} {Experimental quantum teleportation},\ }\href
  {https://doi.org/10.1038/37539} {\bibfield  {journal} {\bibinfo  {journal}
  {Nature}\ }\textbf {\bibinfo {volume} {390}},\ \bibinfo {pages} {575}
  (\bibinfo {year} {1997})}\BibitemShut {NoStop}%
\bibitem [{\citenamefont {\ifmmode~\dot{Z}\else \.{Z}\fi{}ukowski}\ \emph
  {et~al.}(1993)\citenamefont {\ifmmode~\dot{Z}\else \.{Z}\fi{}ukowski},
  \citenamefont {Zeilinger}, \citenamefont {Horne},\ and\ \citenamefont
  {Ekert}}]{Zukowski1993}%
  \BibitemOpen
  \bibfield  {author} {\bibinfo {author} {\bibfnamefont {M.}~\bibnamefont
  {\ifmmode~\dot{Z}\else \.{Z}\fi{}ukowski}}, \bibinfo {author} {\bibfnamefont
  {A.}~\bibnamefont {Zeilinger}}, \bibinfo {author} {\bibfnamefont {M.~A.}\
  \bibnamefont {Horne}},\ and\ \bibinfo {author} {\bibfnamefont {A.~K.}\
  \bibnamefont {Ekert}},\ }\bibfield  {title} {\bibinfo {title}
  {``{E}vent-ready-detectors'' {B}ell experiment via entanglement swapping},\
  }\href {https://doi.org/10.1103/PhysRevLett.71.4287} {\bibfield  {journal}
  {\bibinfo  {journal} {Phys. Rev. Lett.}\ }\textbf {\bibinfo {volume} {71}},\
  \bibinfo {pages} {4287} (\bibinfo {year} {1993})}\BibitemShut {NoStop}%
\bibitem [{\citenamefont {Pan}\ \emph {et~al.}(1998)\citenamefont {Pan},
  \citenamefont {Bouwmeester}, \citenamefont {Weinfurter},\ and\ \citenamefont
  {Zeilinger}}]{Pan1998}%
  \BibitemOpen
  \bibfield  {author} {\bibinfo {author} {\bibfnamefont {J.-W.}\ \bibnamefont
  {Pan}}, \bibinfo {author} {\bibfnamefont {D.}~\bibnamefont {Bouwmeester}},
  \bibinfo {author} {\bibfnamefont {H.}~\bibnamefont {Weinfurter}},\ and\
  \bibinfo {author} {\bibfnamefont {A.}~\bibnamefont {Zeilinger}},\ }\bibfield
  {title} {\bibinfo {title} {Experimental entanglement swapping: Entangling
  photons that never interacted},\ }\href
  {https://doi.org/10.1103/PhysRevLett.80.3891} {\bibfield  {journal} {\bibinfo
   {journal} {Phys. Rev. Lett.}\ }\textbf {\bibinfo {volume} {80}},\ \bibinfo
  {pages} {3891} (\bibinfo {year} {1998})}\BibitemShut {NoStop}%
\bibitem [{\citenamefont {Bennett}\ and\ \citenamefont
  {Wiesner}(1992)}]{Bennett1992}%
  \BibitemOpen
  \bibfield  {author} {\bibinfo {author} {\bibfnamefont {C.~H.}\ \bibnamefont
  {Bennett}}\ and\ \bibinfo {author} {\bibfnamefont {S.~J.}\ \bibnamefont
  {Wiesner}},\ }\bibfield  {title} {\bibinfo {title} {Communication via one-
  and two-particle operators on {Einstein-Podolsky-Rosen} states},\ }\href
  {https://doi.org/10.1103/PhysRevLett.69.2881} {\bibfield  {journal} {\bibinfo
   {journal} {Phys. Rev. Lett.}\ }\textbf {\bibinfo {volume} {69}},\ \bibinfo
  {pages} {2881} (\bibinfo {year} {1992})}\BibitemShut {NoStop}%
\bibitem [{\citenamefont {Mattle}\ \emph {et~al.}(1996)\citenamefont {Mattle},
  \citenamefont {Weinfurter}, \citenamefont {Kwiat},\ and\ \citenamefont
  {Zeilinger}}]{Mattle1996}%
  \BibitemOpen
  \bibfield  {author} {\bibinfo {author} {\bibfnamefont {K.}~\bibnamefont
  {Mattle}}, \bibinfo {author} {\bibfnamefont {H.}~\bibnamefont {Weinfurter}},
  \bibinfo {author} {\bibfnamefont {P.~G.}\ \bibnamefont {Kwiat}},\ and\
  \bibinfo {author} {\bibfnamefont {A.}~\bibnamefont {Zeilinger}},\ }\bibfield
  {title} {\bibinfo {title} {Dense coding in experimental quantum
  communication},\ }\href {https://doi.org/10.1103/PhysRevLett.76.4656}
  {\bibfield  {journal} {\bibinfo  {journal} {Phys. Rev. Lett.}\ }\textbf
  {\bibinfo {volume} {76}},\ \bibinfo {pages} {4656} (\bibinfo {year}
  {1996})}\BibitemShut {NoStop}%
\bibitem [{\citenamefont {Briegel}\ \emph {et~al.}(1998)\citenamefont
  {Briegel}, \citenamefont {D\"ur}, \citenamefont {Cirac},\ and\ \citenamefont
  {Zoller}}]{Briegel1998}%
  \BibitemOpen
  \bibfield  {author} {\bibinfo {author} {\bibfnamefont {H.-J.}\ \bibnamefont
  {Briegel}}, \bibinfo {author} {\bibfnamefont {W.}~\bibnamefont {D\"ur}},
  \bibinfo {author} {\bibfnamefont {J.~I.}\ \bibnamefont {Cirac}},\ and\
  \bibinfo {author} {\bibfnamefont {P.}~\bibnamefont {Zoller}},\ }\bibfield
  {title} {\bibinfo {title} {Quantum repeaters: The role of imperfect local
  operations in quantum communication},\ }\href
  {https://doi.org/10.1103/PhysRevLett.81.5932} {\bibfield  {journal} {\bibinfo
   {journal} {Phys. Rev. Lett.}\ }\textbf {\bibinfo {volume} {81}},\ \bibinfo
  {pages} {5932} (\bibinfo {year} {1998})}\BibitemShut {NoStop}%
\bibitem [{\citenamefont {Kimble}(2008)}]{Kimble2008}%
  \BibitemOpen
  \bibfield  {author} {\bibinfo {author} {\bibfnamefont {H.~J.}\ \bibnamefont
  {Kimble}},\ }\bibfield  {title} {\bibinfo {title} {The quantum internet},\
  }\href {https://doi.org/10.1038/nature07127} {\bibfield  {journal} {\bibinfo
  {journal} {Nature}\ }\textbf {\bibinfo {volume} {453}},\ \bibinfo {pages}
  {1023} (\bibinfo {year} {2008})}\BibitemShut {NoStop}%
\bibitem [{\citenamefont {Wehner}\ \emph {et~al.}(2018)\citenamefont {Wehner},
  \citenamefont {Elkouss},\ and\ \citenamefont {Hanson}}]{Wehner2018}%
  \BibitemOpen
  \bibfield  {author} {\bibinfo {author} {\bibfnamefont {S.}~\bibnamefont
  {Wehner}}, \bibinfo {author} {\bibfnamefont {D.}~\bibnamefont {Elkouss}},\
  and\ \bibinfo {author} {\bibfnamefont {R.}~\bibnamefont {Hanson}},\
  }\bibfield  {title} {\bibinfo {title} {Quantum internet: A vision for the
  road ahead},\ }\href {https://doi.org/10.1126/science.aam9288} {\bibfield
  {journal} {\bibinfo  {journal} {Science}\ }\textbf {\bibinfo {volume}
  {362}},\ \bibinfo {pages} {eaam9288} (\bibinfo {year} {2018})}\BibitemShut
  {NoStop}%
\bibitem [{\citenamefont {Raussendorf}\ and\ \citenamefont
  {Briegel}(2001)}]{Raussendorf2001}%
  \BibitemOpen
  \bibfield  {author} {\bibinfo {author} {\bibfnamefont {R.}~\bibnamefont
  {Raussendorf}}\ and\ \bibinfo {author} {\bibfnamefont {H.~J.}\ \bibnamefont
  {Briegel}},\ }\bibfield  {title} {\bibinfo {title} {A one-way quantum
  computer},\ }\href {https://doi.org/10.1103/PhysRevLett.86.5188} {\bibfield
  {journal} {\bibinfo  {journal} {Phys. Rev. Lett.}\ }\textbf {\bibinfo
  {volume} {86}},\ \bibinfo {pages} {5188} (\bibinfo {year}
  {2001})}\BibitemShut {NoStop}%
\bibitem [{\citenamefont {Knill}\ \emph {et~al.}(2001)\citenamefont {Knill},
  \citenamefont {Laflamme},\ and\ \citenamefont {Milburn}}]{Knill2001}%
  \BibitemOpen
  \bibfield  {author} {\bibinfo {author} {\bibfnamefont {E.}~\bibnamefont
  {Knill}}, \bibinfo {author} {\bibfnamefont {R.}~\bibnamefont {Laflamme}},\
  and\ \bibinfo {author} {\bibfnamefont {G.~J.}\ \bibnamefont {Milburn}},\
  }\bibfield  {title} {\bibinfo {title} {A scheme for efficient quantum
  computation with linear optics},\ }\href {https://doi.org/10.1038/35051009}
  {\bibfield  {journal} {\bibinfo  {journal} {Nature}\ }\textbf {\bibinfo
  {volume} {409}},\ \bibinfo {pages} {46} (\bibinfo {year} {2001})}\BibitemShut
  {NoStop}%
\bibitem [{\citenamefont {Browne}\ and\ \citenamefont
  {Rudolph}(2005)}]{Browne2005}%
  \BibitemOpen
  \bibfield  {author} {\bibinfo {author} {\bibfnamefont {D.~E.}\ \bibnamefont
  {Browne}}\ and\ \bibinfo {author} {\bibfnamefont {T.}~\bibnamefont
  {Rudolph}},\ }\bibfield  {title} {\bibinfo {title} {Resource-efficient linear
  optical quantum computation},\ }\href
  {https://doi.org/10.1103/PhysRevLett.95.010501} {\bibfield  {journal}
  {\bibinfo  {journal} {Phys. Rev. Lett.}\ }\textbf {\bibinfo {volume} {95}},\
  \bibinfo {pages} {010501} (\bibinfo {year} {2005})}\BibitemShut {NoStop}%
\bibitem [{\citenamefont {Kok}\ \emph {et~al.}(2007)\citenamefont {Kok},
  \citenamefont {Munro}, \citenamefont {Nemoto}, \citenamefont {Ralph},
  \citenamefont {Dowling},\ and\ \citenamefont {Milburn}}]{Kok2007}%
  \BibitemOpen
  \bibfield  {author} {\bibinfo {author} {\bibfnamefont {P.}~\bibnamefont
  {Kok}}, \bibinfo {author} {\bibfnamefont {W.~J.}\ \bibnamefont {Munro}},
  \bibinfo {author} {\bibfnamefont {K.}~\bibnamefont {Nemoto}}, \bibinfo
  {author} {\bibfnamefont {T.~C.}\ \bibnamefont {Ralph}}, \bibinfo {author}
  {\bibfnamefont {J.~P.}\ \bibnamefont {Dowling}},\ and\ \bibinfo {author}
  {\bibfnamefont {G.~J.}\ \bibnamefont {Milburn}},\ }\bibfield  {title}
  {\bibinfo {title} {Linear optical quantum computing with photonic qubits},\
  }\href {https://doi.org/10.1103/RevModPhys.79.135} {\bibfield  {journal}
  {\bibinfo  {journal} {Rev. Mod. Phys.}\ }\textbf {\bibinfo {volume} {79}},\
  \bibinfo {pages} {135} (\bibinfo {year} {2007})}\BibitemShut {NoStop}%
\bibitem [{\citenamefont {Weinfurter}(1994)}]{Weinfurter1994}%
  \BibitemOpen
  \bibfield  {author} {\bibinfo {author} {\bibfnamefont {H.}~\bibnamefont
  {Weinfurter}},\ }\bibfield  {title} {\bibinfo {title} {Experimental
  {B}ell-state analysis},\ }\href {https://doi.org/10.1209/0295-5075/25/8/001}
  {\bibfield  {journal} {\bibinfo  {journal} {Europhys. Lett.}\ }\textbf
  {\bibinfo {volume} {25}},\ \bibinfo {pages} {559} (\bibinfo {year}
  {1994})}\BibitemShut {NoStop}%
\bibitem [{\citenamefont {Braunstein}\ and\ \citenamefont
  {Mann}(1995)}]{Braunstein1995}%
  \BibitemOpen
  \bibfield  {author} {\bibinfo {author} {\bibfnamefont {S.~L.}\ \bibnamefont
  {Braunstein}}\ and\ \bibinfo {author} {\bibfnamefont {A.}~\bibnamefont
  {Mann}},\ }\bibfield  {title} {\bibinfo {title} {Measurement of the {B}ell
  operator and quantum teleportation},\ }\href
  {https://doi.org/10.1103/PhysRevA.51.R1727} {\bibfield  {journal} {\bibinfo
  {journal} {Phys. Rev. A}\ }\textbf {\bibinfo {volume} {51}},\ \bibinfo
  {pages} {R1727(R)} (\bibinfo {year} {1995})}\BibitemShut {NoStop}%
\bibitem [{\citenamefont {Michler}\ \emph {et~al.}(1996)\citenamefont
  {Michler}, \citenamefont {Mattle}, \citenamefont {Weinfurter},\ and\
  \citenamefont {Zeilinger}}]{Michler1996}%
  \BibitemOpen
  \bibfield  {author} {\bibinfo {author} {\bibfnamefont {M.}~\bibnamefont
  {Michler}}, \bibinfo {author} {\bibfnamefont {K.}~\bibnamefont {Mattle}},
  \bibinfo {author} {\bibfnamefont {H.}~\bibnamefont {Weinfurter}},\ and\
  \bibinfo {author} {\bibfnamefont {A.}~\bibnamefont {Zeilinger}},\ }\bibfield
  {title} {\bibinfo {title} {Interferometric {B}ell-state analysis},\ }\href
  {https://doi.org/10.1103/PhysRevA.53.R1209} {\bibfield  {journal} {\bibinfo
  {journal} {Phys. Rev. A}\ }\textbf {\bibinfo {volume} {53}},\ \bibinfo
  {pages} {R1209(R)} (\bibinfo {year} {1996})}\BibitemShut {NoStop}%
\bibitem [{\citenamefont {Vaidman}\ and\ \citenamefont
  {Yoran}(1999)}]{Vaidman1999}%
  \BibitemOpen
  \bibfield  {author} {\bibinfo {author} {\bibfnamefont {L.}~\bibnamefont
  {Vaidman}}\ and\ \bibinfo {author} {\bibfnamefont {N.}~\bibnamefont
  {Yoran}},\ }\bibfield  {title} {\bibinfo {title} {Methods for reliable
  teleportation},\ }\href {https://doi.org/10.1103/PhysRevA.59.116} {\bibfield
  {journal} {\bibinfo  {journal} {Phys. Rev. A}\ }\textbf {\bibinfo {volume}
  {59}},\ \bibinfo {pages} {116} (\bibinfo {year} {1999})}\BibitemShut
  {NoStop}%
\bibitem [{\citenamefont {L\"utkenhaus}\ \emph {et~al.}(1999)\citenamefont
  {L\"utkenhaus}, \citenamefont {Calsamiglia},\ and\ \citenamefont
  {Suominen}}]{Lutkenhaus1999}%
  \BibitemOpen
  \bibfield  {author} {\bibinfo {author} {\bibfnamefont {N.}~\bibnamefont
  {L\"utkenhaus}}, \bibinfo {author} {\bibfnamefont {J.}~\bibnamefont
  {Calsamiglia}},\ and\ \bibinfo {author} {\bibfnamefont {K.-A.}\ \bibnamefont
  {Suominen}},\ }\bibfield  {title} {\bibinfo {title} {{B}ell measurements for
  teleportation},\ }\href {https://doi.org/10.1103/PhysRevA.59.3295} {\bibfield
   {journal} {\bibinfo  {journal} {Phys. Rev. A}\ }\textbf {\bibinfo {volume}
  {59}},\ \bibinfo {pages} {3295} (\bibinfo {year} {1999})}\BibitemShut
  {NoStop}%
\bibitem [{\citenamefont {Calsamiglia}\ and\ \citenamefont
  {L{\"u}tkenhaus}(2001)}]{Calsamiglia2001}%
  \BibitemOpen
  \bibfield  {author} {\bibinfo {author} {\bibfnamefont {J.}~\bibnamefont
  {Calsamiglia}}\ and\ \bibinfo {author} {\bibfnamefont {N.}~\bibnamefont
  {L{\"u}tkenhaus}},\ }\bibfield  {title} {\bibinfo {title} {Maximum efficiency
  of a linear-optical {B}ell-state analyzer},\ }\href
  {https://doi.org/10.1007/s003400000484} {\bibfield  {journal} {\bibinfo
  {journal} {Appl. Phys. B}\ }\textbf {\bibinfo {volume} {72}},\ \bibinfo
  {pages} {67} (\bibinfo {year} {2001})}\BibitemShut {NoStop}%
\bibitem [{\citenamefont {Kwiat}\ and\ \citenamefont
  {Weinfurter}(1998)}]{Kwiat1998}%
  \BibitemOpen
  \bibfield  {author} {\bibinfo {author} {\bibfnamefont {P.~G.}\ \bibnamefont
  {Kwiat}}\ and\ \bibinfo {author} {\bibfnamefont {H.}~\bibnamefont
  {Weinfurter}},\ }\bibfield  {title} {\bibinfo {title} {Embedded {B}ell-state
  analysis},\ }\href {https://doi.org/10.1103/PhysRevA.58.R2623} {\bibfield
  {journal} {\bibinfo  {journal} {Phys. Rev. A}\ }\textbf {\bibinfo {volume}
  {58}},\ \bibinfo {pages} {R2623(R)} (\bibinfo {year} {1998})}\BibitemShut
  {NoStop}%
\bibitem [{\citenamefont {Walborn}\ \emph {et~al.}(2003)\citenamefont
  {Walborn}, \citenamefont {P\'adua},\ and\ \citenamefont
  {Monken}}]{Walborn2003}%
  \BibitemOpen
  \bibfield  {author} {\bibinfo {author} {\bibfnamefont {S.~P.}\ \bibnamefont
  {Walborn}}, \bibinfo {author} {\bibfnamefont {S.}~\bibnamefont {P\'adua}},\
  and\ \bibinfo {author} {\bibfnamefont {C.~H.}\ \bibnamefont {Monken}},\
  }\bibfield  {title} {\bibinfo {title} {Hyperentanglement-assisted
  {B}ell-state analysis},\ }\href {https://doi.org/10.1103/PhysRevA.68.042313}
  {\bibfield  {journal} {\bibinfo  {journal} {Phys. Rev. A}\ }\textbf {\bibinfo
  {volume} {68}},\ \bibinfo {pages} {042313} (\bibinfo {year}
  {2003})}\BibitemShut {NoStop}%
\bibitem [{\citenamefont {Schuck}\ \emph {et~al.}(2006)\citenamefont {Schuck},
  \citenamefont {Huber}, \citenamefont {Kurtsiefer},\ and\ \citenamefont
  {Weinfurter}}]{Schuck2006}%
  \BibitemOpen
  \bibfield  {author} {\bibinfo {author} {\bibfnamefont {C.}~\bibnamefont
  {Schuck}}, \bibinfo {author} {\bibfnamefont {G.}~\bibnamefont {Huber}},
  \bibinfo {author} {\bibfnamefont {C.}~\bibnamefont {Kurtsiefer}},\ and\
  \bibinfo {author} {\bibfnamefont {H.}~\bibnamefont {Weinfurter}},\ }\bibfield
   {title} {\bibinfo {title} {Complete deterministic linear optics {B}ell state
  analysis},\ }\href {https://doi.org/10.1103/PhysRevLett.96.190501} {\bibfield
   {journal} {\bibinfo  {journal} {Phys. Rev. Lett.}\ }\textbf {\bibinfo
  {volume} {96}},\ \bibinfo {pages} {190501} (\bibinfo {year}
  {2006})}\BibitemShut {NoStop}%
\bibitem [{\citenamefont {Barbieri}\ \emph {et~al.}(2007)\citenamefont
  {Barbieri}, \citenamefont {Vallone}, \citenamefont {Mataloni},\ and\
  \citenamefont {De~Martini}}]{Barbieri2007}%
  \BibitemOpen
  \bibfield  {author} {\bibinfo {author} {\bibfnamefont {M.}~\bibnamefont
  {Barbieri}}, \bibinfo {author} {\bibfnamefont {G.}~\bibnamefont {Vallone}},
  \bibinfo {author} {\bibfnamefont {P.}~\bibnamefont {Mataloni}},\ and\
  \bibinfo {author} {\bibfnamefont {F.}~\bibnamefont {De~Martini}},\ }\bibfield
   {title} {\bibinfo {title} {Complete and deterministic discrimination of
  polarization {B}ell states assisted by momentum entanglement},\ }\href
  {https://doi.org/10.1103/PhysRevA.75.042317} {\bibfield  {journal} {\bibinfo
  {journal} {Phys. Rev. A}\ }\textbf {\bibinfo {volume} {75}},\ \bibinfo
  {pages} {042317} (\bibinfo {year} {2007})}\BibitemShut {NoStop}%
\bibitem [{\citenamefont {Wei}\ \emph {et~al.}(2007)\citenamefont {Wei},
  \citenamefont {Barreiro},\ and\ \citenamefont {Kwiat}}]{Wei2007}%
  \BibitemOpen
  \bibfield  {author} {\bibinfo {author} {\bibfnamefont {T.-C.}\ \bibnamefont
  {Wei}}, \bibinfo {author} {\bibfnamefont {J.~T.}\ \bibnamefont {Barreiro}},\
  and\ \bibinfo {author} {\bibfnamefont {P.~G.}\ \bibnamefont {Kwiat}},\
  }\bibfield  {title} {\bibinfo {title} {Hyperentangled {B}ell-state
  analysis},\ }\href {https://doi.org/10.1103/PhysRevA.75.060305} {\bibfield
  {journal} {\bibinfo  {journal} {Phys. Rev. A}\ }\textbf {\bibinfo {volume}
  {75}},\ \bibinfo {pages} {060305(R)} (\bibinfo {year} {2007})}\BibitemShut
  {NoStop}%
\bibitem [{\citenamefont {Grice}(2011)}]{Grice2011}%
  \BibitemOpen
  \bibfield  {author} {\bibinfo {author} {\bibfnamefont {W.~P.}\ \bibnamefont
  {Grice}},\ }\bibfield  {title} {\bibinfo {title} {Arbitrarily complete
  {B}ell-state measurement using only linear optical elements},\ }\href
  {https://doi.org/10.1103/PhysRevA.84.042331} {\bibfield  {journal} {\bibinfo
  {journal} {Phys. Rev. A}\ }\textbf {\bibinfo {volume} {84}},\ \bibinfo
  {pages} {042331} (\bibinfo {year} {2011})}\BibitemShut {NoStop}%
\bibitem [{\citenamefont {Ewert}\ and\ \citenamefont {van
  Loock}(2014)}]{Ewert2014}%
  \BibitemOpen
  \bibfield  {author} {\bibinfo {author} {\bibfnamefont {F.}~\bibnamefont
  {Ewert}}\ and\ \bibinfo {author} {\bibfnamefont {P.}~\bibnamefont {van
  Loock}},\ }\bibfield  {title} {\bibinfo {title} {$3/4$-efficient {B}ell
  measurement with passive linear optics and unentangled ancillae},\ }\href
  {https://doi.org/10.1103/PhysRevLett.113.140403} {\bibfield  {journal}
  {\bibinfo  {journal} {Phys. Rev. Lett.}\ }\textbf {\bibinfo {volume} {113}},\
  \bibinfo {pages} {140403} (\bibinfo {year} {2014})}\BibitemShut {NoStop}%
\bibitem [{\citenamefont {Olivo}\ and\ \citenamefont
  {Grosshans}(2018)}]{Olivo2018}%
  \BibitemOpen
  \bibfield  {author} {\bibinfo {author} {\bibfnamefont {A.}~\bibnamefont
  {Olivo}}\ and\ \bibinfo {author} {\bibfnamefont {F.}~\bibnamefont
  {Grosshans}},\ }\bibfield  {title} {\bibinfo {title} {Ancilla-assisted linear
  optical {B}ell measurements and their optimality},\ }\href
  {https://doi.org/10.1103/PhysRevA.98.042323} {\bibfield  {journal} {\bibinfo
  {journal} {Phys. Rev. A}\ }\textbf {\bibinfo {volume} {98}},\ \bibinfo
  {pages} {042323} (\bibinfo {year} {2018})}\BibitemShut {NoStop}%
\bibitem [{\citenamefont {Bayerbach}\ \emph {et~al.}(2023)\citenamefont
  {Bayerbach}, \citenamefont {D’Aurelio}, \citenamefont {van Loock},\ and\
  \citenamefont {Barz}}]{Bayerbach2023}%
  \BibitemOpen
  \bibfield  {author} {\bibinfo {author} {\bibfnamefont {M.~J.}\ \bibnamefont
  {Bayerbach}}, \bibinfo {author} {\bibfnamefont {S.~E.}\ \bibnamefont
  {D’Aurelio}}, \bibinfo {author} {\bibfnamefont {P.}~\bibnamefont {van
  Loock}},\ and\ \bibinfo {author} {\bibfnamefont {S.}~\bibnamefont {Barz}},\
  }\bibfield  {title} {\bibinfo {title} {{B}ell-state measurement exceeding
  50\% success probability with linear optics},\ }\href
  {https://doi.org/10.1126/sciadv.adf4080} {\bibfield  {journal} {\bibinfo
  {journal} {Sci. Adv.}\ }\textbf {\bibinfo {volume} {9}},\ \bibinfo {pages}
  {eadf4080} (\bibinfo {year} {2023})}\BibitemShut {NoStop}%
\bibitem [{\citenamefont {Hauser}\ \emph {et~al.}(2025)\citenamefont {Hauser},
  \citenamefont {Bayerbach}, \citenamefont {D'Aurelio}, \citenamefont {Weber},
  \citenamefont {Santandrea}, \citenamefont {Kumar}, \citenamefont {Dhand},\
  and\ \citenamefont {Barz}}]{Hauser2025}%
  \BibitemOpen
  \bibfield  {author} {\bibinfo {author} {\bibfnamefont {N.}~\bibnamefont
  {Hauser}}, \bibinfo {author} {\bibfnamefont {M.~J.}\ \bibnamefont
  {Bayerbach}}, \bibinfo {author} {\bibfnamefont {S.~E.}\ \bibnamefont
  {D'Aurelio}}, \bibinfo {author} {\bibfnamefont {R.}~\bibnamefont {Weber}},
  \bibinfo {author} {\bibfnamefont {M.}~\bibnamefont {Santandrea}}, \bibinfo
  {author} {\bibfnamefont {S.~P.}\ \bibnamefont {Kumar}}, \bibinfo {author}
  {\bibfnamefont {I.}~\bibnamefont {Dhand}},\ and\ \bibinfo {author}
  {\bibfnamefont {S.}~\bibnamefont {Barz}},\ }\bibfield  {title} {\bibinfo
  {title} {Boosted bell-state measurements for photonic quantum computation},\
  }\href {https://doi.org/10.1038/s41534-025-00986-2} {\bibfield  {journal}
  {\bibinfo  {journal} {npj Quantum Inf.}\ }\textbf {\bibinfo {volume} {11}},\
  \bibinfo {pages} {41} (\bibinfo {year} {2025})}\BibitemShut {NoStop}%
\bibitem [{\citenamefont {Baghdasaryan}\ \emph {et~al.}(2025)\citenamefont
  {Baghdasaryan}, \citenamefont {Joarder},\ and\ \citenamefont
  {Steinlechner}}]{Baghdasar2025}%
  \BibitemOpen
  \bibfield  {author} {\bibinfo {author} {\bibfnamefont {B.}~\bibnamefont
  {Baghdasaryan}}, \bibinfo {author} {\bibfnamefont {K.}~\bibnamefont
  {Joarder}},\ and\ \bibinfo {author} {\bibfnamefont {F.}~\bibnamefont
  {Steinlechner}},\ }\href {https://arxiv.org/abs/2509.02817} {\bibinfo {title}
  {Efficient entanglement swapping in high-dimensions with only linear optics}}
  (\bibinfo {year} {2025}),\ \Eprint {https://arxiv.org/abs/2509.02817}
  {arXiv:2509.02817 [quant-ph]} \BibitemShut {NoStop}%
\bibitem [{\citenamefont {Zaidi}\ and\ \citenamefont {van
  Loock}(2013)}]{Zaidi2013}%
  \BibitemOpen
  \bibfield  {author} {\bibinfo {author} {\bibfnamefont {H.~A.}\ \bibnamefont
  {Zaidi}}\ and\ \bibinfo {author} {\bibfnamefont {P.}~\bibnamefont {van
  Loock}},\ }\bibfield  {title} {\bibinfo {title} {Beating the one-half limit
  of ancilla-free linear optics {B}ell measurements},\ }\href
  {https://doi.org/10.1103/PhysRevLett.110.260501} {\bibfield  {journal}
  {\bibinfo  {journal} {Phys. Rev. Lett.}\ }\textbf {\bibinfo {volume} {110}},\
  \bibinfo {pages} {260501} (\bibinfo {year} {2013})}\BibitemShut {NoStop}%
\bibitem [{\citenamefont {Kilmer}\ and\ \citenamefont
  {Guha}(2019)}]{Kilmer2019}%
  \BibitemOpen
  \bibfield  {author} {\bibinfo {author} {\bibfnamefont {T.}~\bibnamefont
  {Kilmer}}\ and\ \bibinfo {author} {\bibfnamefont {S.}~\bibnamefont {Guha}},\
  }\bibfield  {title} {\bibinfo {title} {Boosting linear-optical {B}ell
  measurement success probability with predetection squeezing and imperfect
  photon-number-resolving detectors},\ }\href
  {https://doi.org/10.1103/PhysRevA.99.032302} {\bibfield  {journal} {\bibinfo
  {journal} {Phys. Rev. A}\ }\textbf {\bibinfo {volume} {99}},\ \bibinfo
  {pages} {032302} (\bibinfo {year} {2019})}\BibitemShut {NoStop}%
\bibitem [{\citenamefont {Bianchi}\ \emph
  {et~al.}(2025{\natexlab{a}})\citenamefont {Bianchi}, \citenamefont {Marconi},
  \citenamefont {Sperling},\ and\ \citenamefont {Bacco}}]{Bianchi2025}%
  \BibitemOpen
  \bibfield  {author} {\bibinfo {author} {\bibfnamefont {L.}~\bibnamefont
  {Bianchi}}, \bibinfo {author} {\bibfnamefont {C.}~\bibnamefont {Marconi}},
  \bibinfo {author} {\bibfnamefont {J.}~\bibnamefont {Sperling}},\ and\
  \bibinfo {author} {\bibfnamefont {D.}~\bibnamefont {Bacco}},\ }\bibfield
  {title} {\bibinfo {title} {Predetection squeezing as a resource for
  high-dimensional {B}ell-state measurements},\ }\href
  {https://doi.org/10.1103/PhysRevResearch.7.023038} {\bibfield  {journal}
  {\bibinfo  {journal} {Phys. Rev. Res.}\ }\textbf {\bibinfo {volume} {7}},\
  \bibinfo {pages} {023038} (\bibinfo {year} {2025}{\natexlab{a}})}\BibitemShut
  {NoStop}%
\bibitem [{\citenamefont {Bianchi}\ \emph
  {et~al.}(2025{\natexlab{b}})\citenamefont {Bianchi}, \citenamefont {Marconi},
  \citenamefont {Guarda},\ and\ \citenamefont {Bacco}}]{Bianchi2025_PRA}%
  \BibitemOpen
  \bibfield  {author} {\bibinfo {author} {\bibfnamefont {L.}~\bibnamefont
  {Bianchi}}, \bibinfo {author} {\bibfnamefont {C.}~\bibnamefont {Marconi}},
  \bibinfo {author} {\bibfnamefont {G.}~\bibnamefont {Guarda}},\ and\ \bibinfo
  {author} {\bibfnamefont {D.}~\bibnamefont {Bacco}},\ }\bibfield  {title}
  {\bibinfo {title} {Nonlinear protocol for high-dimensional quantum
  teleportation},\ }\href {https://doi.org/10.1103/3x44-664w} {\bibfield
  {journal} {\bibinfo  {journal} {Phys. Rev. A}\ }\textbf {\bibinfo {volume}
  {112}},\ \bibinfo {pages} {012615} (\bibinfo {year}
  {2025}{\natexlab{b}})}\BibitemShut {NoStop}%
\bibitem [{\citenamefont {D'Aurelio}\ \emph {et~al.}(2025)\citenamefont
  {D'Aurelio}, \citenamefont {Bayerbach},\ and\ \citenamefont
  {Barz}}]{DAurelio2025}%
  \BibitemOpen
  \bibfield  {author} {\bibinfo {author} {\bibfnamefont {S.~E.}\ \bibnamefont
  {D'Aurelio}}, \bibinfo {author} {\bibfnamefont {M.~J.}\ \bibnamefont
  {Bayerbach}},\ and\ \bibinfo {author} {\bibfnamefont {S.}~\bibnamefont
  {Barz}},\ }\bibfield  {title} {\bibinfo {title} {Boosted quantum
  teleportation},\ }\href {https://doi.org/10.1038/s41534-025-00992-4}
  {\bibfield  {journal} {\bibinfo  {journal} {npj Quantum Inf.}\ }\textbf
  {\bibinfo {volume} {11}},\ \bibinfo {pages} {37} (\bibinfo {year}
  {2025})}\BibitemShut {NoStop}%
\bibitem [{\citenamefont {Bacco}\ \emph {et~al.}(2021)\citenamefont {Bacco},
  \citenamefont {Bulmer}, \citenamefont {Erhard}, \citenamefont {Huber},\ and\
  \citenamefont {Paesani}}]{Bacco2021}%
  \BibitemOpen
  \bibfield  {author} {\bibinfo {author} {\bibfnamefont {D.}~\bibnamefont
  {Bacco}}, \bibinfo {author} {\bibfnamefont {J.~F.~F.}\ \bibnamefont
  {Bulmer}}, \bibinfo {author} {\bibfnamefont {M.}~\bibnamefont {Erhard}},
  \bibinfo {author} {\bibfnamefont {M.}~\bibnamefont {Huber}},\ and\ \bibinfo
  {author} {\bibfnamefont {S.}~\bibnamefont {Paesani}},\ }\bibfield  {title}
  {\bibinfo {title} {Proposal for practical multidimensional quantum
  networks},\ }\href {https://doi.org/10.1103/PhysRevA.104.052618} {\bibfield
  {journal} {\bibinfo  {journal} {Phys. Rev. A}\ }\textbf {\bibinfo {volume}
  {104}},\ \bibinfo {pages} {052618} (\bibinfo {year} {2021})}\BibitemShut
  {NoStop}%
\bibitem [{\citenamefont {Mirhosseini}\ \emph {et~al.}(2015)\citenamefont
  {Mirhosseini}, \citenamefont {Magaña-Loaiza}, \citenamefont {O’Sullivan},
  \citenamefont {Rodenburg}, \citenamefont {Malik}, \citenamefont {Lavery},
  \citenamefont {Padgett}, \citenamefont {Gauthier},\ and\ \citenamefont
  {Boyd}}]{Mirhosseini2015}%
  \BibitemOpen
  \bibfield  {author} {\bibinfo {author} {\bibfnamefont {M.}~\bibnamefont
  {Mirhosseini}}, \bibinfo {author} {\bibfnamefont {O.~S.}\ \bibnamefont
  {Magaña-Loaiza}}, \bibinfo {author} {\bibfnamefont {M.~N.}\ \bibnamefont
  {O’Sullivan}}, \bibinfo {author} {\bibfnamefont {B.}~\bibnamefont
  {Rodenburg}}, \bibinfo {author} {\bibfnamefont {M.}~\bibnamefont {Malik}},
  \bibinfo {author} {\bibfnamefont {M.~P.~J.}\ \bibnamefont {Lavery}}, \bibinfo
  {author} {\bibfnamefont {M.~J.}\ \bibnamefont {Padgett}}, \bibinfo {author}
  {\bibfnamefont {D.~J.}\ \bibnamefont {Gauthier}},\ and\ \bibinfo {author}
  {\bibfnamefont {R.~W.}\ \bibnamefont {Boyd}},\ }\bibfield  {title} {\bibinfo
  {title} {High-dimensional quantum cryptography with twisted light},\ }\href
  {https://doi.org/10.1088/1367-2630/17/3/033033} {\bibfield  {journal}
  {\bibinfo  {journal} {New J. Phys.}\ }\textbf {\bibinfo {volume} {17}},\
  \bibinfo {pages} {033033} (\bibinfo {year} {2015})}\BibitemShut {NoStop}%
\bibitem [{\citenamefont {Nape}\ \emph {et~al.}(2023)\citenamefont {Nape},
  \citenamefont {Sephton}, \citenamefont {Ornelas}, \citenamefont {Moodley},\
  and\ \citenamefont {Forbes}}]{Nape2023}%
  \BibitemOpen
  \bibfield  {author} {\bibinfo {author} {\bibfnamefont {I.}~\bibnamefont
  {Nape}}, \bibinfo {author} {\bibfnamefont {B.}~\bibnamefont {Sephton}},
  \bibinfo {author} {\bibfnamefont {P.}~\bibnamefont {Ornelas}}, \bibinfo
  {author} {\bibfnamefont {C.}~\bibnamefont {Moodley}},\ and\ \bibinfo {author}
  {\bibfnamefont {A.}~\bibnamefont {Forbes}},\ }\bibfield  {title} {\bibinfo
  {title} {Quantum structured light in high dimensions},\ }\href
  {https://doi.org/10.1063/5.0138224} {\bibfield  {journal} {\bibinfo
  {journal} {APL Photonics}\ }\textbf {\bibinfo {volume} {8}},\ \bibinfo
  {pages} {051101} (\bibinfo {year} {2023})}\BibitemShut {NoStop}%
\bibitem [{\citenamefont {Cozzolino}\ \emph {et~al.}(2019)\citenamefont
  {Cozzolino}, \citenamefont {Da~Lio}, \citenamefont {Bacco},\ and\
  \citenamefont {Oxenløwe}}]{Cozzolino2019}%
  \BibitemOpen
  \bibfield  {author} {\bibinfo {author} {\bibfnamefont {D.}~\bibnamefont
  {Cozzolino}}, \bibinfo {author} {\bibfnamefont {B.}~\bibnamefont {Da~Lio}},
  \bibinfo {author} {\bibfnamefont {D.}~\bibnamefont {Bacco}},\ and\ \bibinfo
  {author} {\bibfnamefont {L.~K.}\ \bibnamefont {Oxenløwe}},\ }\bibfield
  {title} {\bibinfo {title} {High-dimensional quantum communication: Benefits,
  progress, and future challenges},\ }\href
  {https://doi.org/https://doi.org/10.1002/qute.201900038} {\bibfield
  {journal} {\bibinfo  {journal} {Adv. Quantum Technol.}\ }\textbf {\bibinfo
  {volume} {2}},\ \bibinfo {pages} {1900038} (\bibinfo {year}
  {2019})}\BibitemShut {NoStop}%
\bibitem [{\citenamefont {Erhard}\ \emph {et~al.}(2020)\citenamefont {Erhard},
  \citenamefont {Krenn},\ and\ \citenamefont {Zeilinger}}]{Erhard2020}%
  \BibitemOpen
  \bibfield  {author} {\bibinfo {author} {\bibfnamefont {M.}~\bibnamefont
  {Erhard}}, \bibinfo {author} {\bibfnamefont {M.}~\bibnamefont {Krenn}},\ and\
  \bibinfo {author} {\bibfnamefont {A.}~\bibnamefont {Zeilinger}},\ }\bibfield
  {title} {\bibinfo {title} {Advances in high-dimensional quantum
  entanglement},\ }\href {https://doi.org/10.1038/s42254-020-0193-5} {\bibfield
   {journal} {\bibinfo  {journal} {Nat. Rev. Phys.}\ }\textbf {\bibinfo
  {volume} {2}},\ \bibinfo {pages} {365} (\bibinfo {year} {2020})}\BibitemShut
  {NoStop}%
\bibitem [{\citenamefont {Calsamiglia}(2002)}]{Calsamiglia2002}%
  \BibitemOpen
  \bibfield  {author} {\bibinfo {author} {\bibfnamefont {J.}~\bibnamefont
  {Calsamiglia}},\ }\bibfield  {title} {\bibinfo {title} {Generalized
  measurements by linear elements},\ }\href
  {https://doi.org/10.1103/PhysRevA.65.030301} {\bibfield  {journal} {\bibinfo
  {journal} {Phys. Rev. A}\ }\textbf {\bibinfo {volume} {65}},\ \bibinfo
  {pages} {030301(R)} (\bibinfo {year} {2002})}\BibitemShut {NoStop}%
\bibitem [{\citenamefont {Dušek}(2001)}]{Dusek2001}%
  \BibitemOpen
  \bibfield  {author} {\bibinfo {author} {\bibfnamefont {M.}~\bibnamefont
  {Dušek}},\ }\bibfield  {title} {\bibinfo {title} {Discrimination of the
  {B}ell states of qudits by means of linear optics},\ }\href
  {https://doi.org/https://doi.org/10.1016/S0030-4018(01)01565-6} {\bibfield
  {journal} {\bibinfo  {journal} {Optics Commun.}\ }\textbf {\bibinfo {volume}
  {199}},\ \bibinfo {pages} {161} (\bibinfo {year} {2001})}\BibitemShut
  {NoStop}%
\bibitem [{\citenamefont {Zhang}\ \emph
  {et~al.}(2019{\natexlab{a}})\citenamefont {Zhang}, \citenamefont {Zhang},
  \citenamefont {Hu}, \citenamefont {Liu}, \citenamefont {Huang}, \citenamefont
  {Li},\ and\ \citenamefont {Guo}}]{Zhang2019}%
  \BibitemOpen
  \bibfield  {author} {\bibinfo {author} {\bibfnamefont {H.}~\bibnamefont
  {Zhang}}, \bibinfo {author} {\bibfnamefont {C.}~\bibnamefont {Zhang}},
  \bibinfo {author} {\bibfnamefont {X.-M.}\ \bibnamefont {Hu}}, \bibinfo
  {author} {\bibfnamefont {B.-H.}\ \bibnamefont {Liu}}, \bibinfo {author}
  {\bibfnamefont {Y.-F.}\ \bibnamefont {Huang}}, \bibinfo {author}
  {\bibfnamefont {C.-F.}\ \bibnamefont {Li}},\ and\ \bibinfo {author}
  {\bibfnamefont {G.-C.}\ \bibnamefont {Guo}},\ }\bibfield  {title} {\bibinfo
  {title} {Arbitrary two-particle high-dimensional {B}ell-state measurement by
  auxiliary entanglement},\ }\href {https://doi.org/10.1103/PhysRevA.99.052301}
  {\bibfield  {journal} {\bibinfo  {journal} {Phys. Rev. A}\ }\textbf {\bibinfo
  {volume} {99}},\ \bibinfo {pages} {052301} (\bibinfo {year}
  {2019}{\natexlab{a}})}\BibitemShut {NoStop}%
\bibitem [{\citenamefont {Bharos}\ \emph {et~al.}(2025)\citenamefont {Bharos},
  \citenamefont {Markovich},\ and\ \citenamefont {Borregaard}}]{Bharos2025}%
  \BibitemOpen
  \bibfield  {author} {\bibinfo {author} {\bibfnamefont {N.}~\bibnamefont
  {Bharos}}, \bibinfo {author} {\bibfnamefont {L.}~\bibnamefont {Markovich}},\
  and\ \bibinfo {author} {\bibfnamefont {J.}~\bibnamefont {Borregaard}},\
  }\bibfield  {title} {\bibinfo {title} {Efficient high-dimensional entangled
  state analyzer with linear optics},\ }\href
  {https://doi.org/10.22331/q-2025-04-18-1711} {\bibfield  {journal} {\bibinfo
  {journal} {Quantum}\ }\textbf {\bibinfo {volume} {9}},\ \bibinfo {pages}
  {1711} (\bibinfo {year} {2025})}\BibitemShut {NoStop}%
\bibitem [{\citenamefont {Lee}\ \emph {et~al.}(2015)\citenamefont {Lee},
  \citenamefont {Park}, \citenamefont {Ralph},\ and\ \citenamefont
  {Jeong}}]{Lee_PRL_2015}%
  \BibitemOpen
  \bibfield  {author} {\bibinfo {author} {\bibfnamefont {S.-W.}\ \bibnamefont
  {Lee}}, \bibinfo {author} {\bibfnamefont {K.}~\bibnamefont {Park}}, \bibinfo
  {author} {\bibfnamefont {T.~C.}\ \bibnamefont {Ralph}},\ and\ \bibinfo
  {author} {\bibfnamefont {H.}~\bibnamefont {Jeong}},\ }\bibfield  {title}
  {\bibinfo {title} {Nearly deterministic bell measurement for multiphoton
  qubits and its application to quantum information processing},\ }\href
  {https://doi.org/10.1103/PhysRevLett.114.113603} {\bibfield  {journal}
  {\bibinfo  {journal} {Phys. Rev. Lett.}\ }\textbf {\bibinfo {volume} {114}},\
  \bibinfo {pages} {113603} (\bibinfo {year} {2015})}\BibitemShut {NoStop}%
\bibitem [{\citenamefont {Ewert}\ \emph {et~al.}(2016)\citenamefont {Ewert},
  \citenamefont {Bergmann},\ and\ \citenamefont {van Loock}}]{Ewert_PRL_2016}%
  \BibitemOpen
  \bibfield  {author} {\bibinfo {author} {\bibfnamefont {F.}~\bibnamefont
  {Ewert}}, \bibinfo {author} {\bibfnamefont {M.}~\bibnamefont {Bergmann}},\
  and\ \bibinfo {author} {\bibfnamefont {P.}~\bibnamefont {van Loock}},\
  }\bibfield  {title} {\bibinfo {title} {Ultrafast long-distance quantum
  communication with static linear optics},\ }\href
  {https://doi.org/10.1103/PhysRevLett.117.210501} {\bibfield  {journal}
  {\bibinfo  {journal} {Phys. Rev. Lett.}\ }\textbf {\bibinfo {volume} {117}},\
  \bibinfo {pages} {210501} (\bibinfo {year} {2016})}\BibitemShut {NoStop}%
\bibitem [{\citenamefont {Ewert}\ and\ \citenamefont {van
  Loock}(2017)}]{Ewert_PRA_2017}%
  \BibitemOpen
  \bibfield  {author} {\bibinfo {author} {\bibfnamefont {F.}~\bibnamefont
  {Ewert}}\ and\ \bibinfo {author} {\bibfnamefont {P.}~\bibnamefont {van
  Loock}},\ }\bibfield  {title} {\bibinfo {title} {Ultrafast fault-tolerant
  long-distance quantum communication with static linear optics},\ }\href
  {https://doi.org/10.1103/PhysRevA.95.012327} {\bibfield  {journal} {\bibinfo
  {journal} {Phys. Rev. A}\ }\textbf {\bibinfo {volume} {95}},\ \bibinfo
  {pages} {012327} (\bibinfo {year} {2017})}\BibitemShut {NoStop}%
\bibitem [{\citenamefont {Lee}\ \emph {et~al.}(2019)\citenamefont {Lee},
  \citenamefont {Ralph},\ and\ \citenamefont {Jeong}}]{Lee_PRA_2019}%
  \BibitemOpen
  \bibfield  {author} {\bibinfo {author} {\bibfnamefont {S.-W.}\ \bibnamefont
  {Lee}}, \bibinfo {author} {\bibfnamefont {T.~C.}\ \bibnamefont {Ralph}},\
  and\ \bibinfo {author} {\bibfnamefont {H.}~\bibnamefont {Jeong}},\ }\bibfield
   {title} {\bibinfo {title} {Fundamental building block for all-optical
  scalable quantum networks},\ }\href
  {https://doi.org/10.1103/PhysRevA.100.052303} {\bibfield  {journal} {\bibinfo
   {journal} {Phys. Rev. A}\ }\textbf {\bibinfo {volume} {100}},\ \bibinfo
  {pages} {052303} (\bibinfo {year} {2019})}\BibitemShut {NoStop}%
\bibitem [{\citenamefont {Schmidt}\ and\ \citenamefont {van
  Loock}(2019)}]{Schmidt_PRA_2019}%
  \BibitemOpen
  \bibfield  {author} {\bibinfo {author} {\bibfnamefont {F.}~\bibnamefont
  {Schmidt}}\ and\ \bibinfo {author} {\bibfnamefont {P.}~\bibnamefont {van
  Loock}},\ }\bibfield  {title} {\bibinfo {title} {Efficiencies of logical bell
  measurements on calderbank-shor-steane codes with static linear optics},\
  }\href {https://doi.org/10.1103/PhysRevA.99.062308} {\bibfield  {journal}
  {\bibinfo  {journal} {Phys. Rev. A}\ }\textbf {\bibinfo {volume} {99}},\
  \bibinfo {pages} {062308} (\bibinfo {year} {2019})}\BibitemShut {NoStop}%
\bibitem [{\citenamefont {Hilaire}\ \emph {et~al.}(2023)\citenamefont
  {Hilaire}, \citenamefont {Castor}, \citenamefont {Barnes}, \citenamefont
  {Economou},\ and\ \citenamefont {Grosshans}}]{Hilaire2023}%
  \BibitemOpen
  \bibfield  {author} {\bibinfo {author} {\bibfnamefont {P.}~\bibnamefont
  {Hilaire}}, \bibinfo {author} {\bibfnamefont {Y.}~\bibnamefont {Castor}},
  \bibinfo {author} {\bibfnamefont {E.}~\bibnamefont {Barnes}}, \bibinfo
  {author} {\bibfnamefont {S.~E.}\ \bibnamefont {Economou}},\ and\ \bibinfo
  {author} {\bibfnamefont {F.}~\bibnamefont {Grosshans}},\ }\bibfield  {title}
  {\bibinfo {title} {Linear optical logical {B}ell state measurements with
  optimal loss-tolerance threshold},\ }\href
  {https://doi.org/10.1103/PRXQuantum.4.040322} {\bibfield  {journal} {\bibinfo
   {journal} {PRX Quantum}\ }\textbf {\bibinfo {volume} {4}},\ \bibinfo {pages}
  {040322} (\bibinfo {year} {2023})}\BibitemShut {NoStop}%
\bibitem [{\citenamefont {Bell}\ \emph {et~al.}(2023)\citenamefont {Bell},
  \citenamefont {Pettersson},\ and\ \citenamefont
  {Paesani}}]{Bell_PRXQuantum_2023}%
  \BibitemOpen
  \bibfield  {author} {\bibinfo {author} {\bibfnamefont {T.~J.}\ \bibnamefont
  {Bell}}, \bibinfo {author} {\bibfnamefont {L.~A.}\ \bibnamefont
  {Pettersson}},\ and\ \bibinfo {author} {\bibfnamefont {S.}~\bibnamefont
  {Paesani}},\ }\bibfield  {title} {\bibinfo {title} {Optimizing graph codes
  for measurement-based loss tolerance},\ }\href
  {https://doi.org/10.1103/PRXQuantum.4.020328} {\bibfield  {journal} {\bibinfo
   {journal} {PRX Quantum}\ }\textbf {\bibinfo {volume} {4}},\ \bibinfo {pages}
  {020328} (\bibinfo {year} {2023})}\BibitemShut {NoStop}%
\bibitem [{\citenamefont {Schmidt}\ and\ \citenamefont {van
  Loock}(2024)}]{Schmidt2024Fusion}%
  \BibitemOpen
  \bibfield  {author} {\bibinfo {author} {\bibfnamefont {F.}~\bibnamefont
  {Schmidt}}\ and\ \bibinfo {author} {\bibfnamefont {P.}~\bibnamefont {van
  Loock}},\ }\href {https://arxiv.org/abs/2410.20261} {\bibinfo {title}
  {Generalized fusions of photonic quantum states using static linear optics}}
  (\bibinfo {year} {2024}),\ \Eprint {https://arxiv.org/abs/2410.20261}
  {arXiv:2410.20261 [quant-ph]} \BibitemShut {NoStop}%
\bibitem [{\citenamefont {\"Ust\"un}\ \emph {et~al.}(2025)\citenamefont
  {\"Ust\"un}, \citenamefont {Rieffel}, \citenamefont {Devitt},\ and\
  \citenamefont {Saied}}]{Ustun2025}%
  \BibitemOpen
  \bibfield  {author} {\bibinfo {author} {\bibfnamefont {G.}~\bibnamefont
  {\"Ust\"un}}, \bibinfo {author} {\bibfnamefont {E.~G.}\ \bibnamefont
  {Rieffel}}, \bibinfo {author} {\bibfnamefont {S.~J.}\ \bibnamefont
  {Devitt}},\ and\ \bibinfo {author} {\bibfnamefont {J.}~\bibnamefont
  {Saied}},\ }\bibfield  {title} {\bibinfo {title} {Fusion for high-dimensional
  linear-optical quantum computing with improved success probability},\ }\href
  {https://doi.org/10.1103/l7bg-hc8c} {\bibfield  {journal} {\bibinfo
  {journal} {Phys. Rev. Appl.}\ }\textbf {\bibinfo {volume} {24}},\ \bibinfo
  {pages} {044024} (\bibinfo {year} {2025})}\BibitemShut {NoStop}%
\bibitem [{\citenamefont {Yamazaki}\ and\ \citenamefont
  {Azuma}(2025)}]{Yamazaki2025}%
  \BibitemOpen
  \bibfield  {author} {\bibinfo {author} {\bibfnamefont {T.}~\bibnamefont
  {Yamazaki}}\ and\ \bibinfo {author} {\bibfnamefont {K.}~\bibnamefont
  {Azuma}},\ }\bibfield  {title} {\bibinfo {title} {Linear-optical fusion
  boosted by high-dimensional entanglement},\ }\href
  {https://doi.org/10.1103/PhysRevLett.134.200801} {\bibfield  {journal}
  {\bibinfo  {journal} {Phys. Rev. Lett.}\ }\textbf {\bibinfo {volume} {134}},\
  \bibinfo {pages} {200801} (\bibinfo {year} {2025})}\BibitemShut {NoStop}%
\bibitem [{\citenamefont {Reiß}\ and\ \citenamefont {van
  Loock}(2026)}]{Reiss2026LogicalBM}%
  \BibitemOpen
  \bibfield  {author} {\bibinfo {author} {\bibfnamefont {S.~D.}\ \bibnamefont
  {Reiß}}\ and\ \bibinfo {author} {\bibfnamefont {P.}~\bibnamefont {van
  Loock}},\ }\href {https://arxiv.org/abs/2601.08820} {\bibinfo {title}
  {Optimal logical {B}ell measurements on stabilizer codes with linear optics}}
  (\bibinfo {year} {2026}),\ \Eprint {https://arxiv.org/abs/2601.08820}
  {arXiv:2601.08820 [quant-ph]} \BibitemShut {NoStop}%
\bibitem [{\citenamefont {Laha}\ and\ \citenamefont {van
  Loock}(2026)}]{Laha_HDBM_Sqz_2026}%
  \BibitemOpen
  \bibfield  {author} {\bibinfo {author} {\bibfnamefont {P.}~\bibnamefont
  {Laha}}\ and\ \bibinfo {author} {\bibfnamefont {P.}~\bibnamefont {van
  Loock}},\ }\href {https://arxiv.org/abs/2606.29432} {\bibinfo {title}
  {Squeezing-enhanced pairwise fusion of photonic qudits}} (\bibinfo {year}
  {2026}),\ \Eprint {https://arxiv.org/abs/2606.29432} {arXiv:2606.29432
  [quant-ph]} \BibitemShut {NoStop}%
\bibitem [{\citenamefont {Zeng}(2025)}]{Zeng2025}%
  \BibitemOpen
  \bibfield  {author} {\bibinfo {author} {\bibfnamefont {Z.}~\bibnamefont
  {Zeng}},\ }\bibfield  {title} {\bibinfo {title} {Linear-optical
  four-dimensional {B}ell state measurement for superdense coding assisted by
  polarization entanglement},\ }\href {https://doi.org/10.1364/JOSAB.539895}
  {\bibfield  {journal} {\bibinfo  {journal} {J. Opt. Soc. Am. B}\ }\textbf
  {\bibinfo {volume} {42}},\ \bibinfo {pages} {93} (\bibinfo {year}
  {2025})}\BibitemShut {NoStop}%
\bibitem [{\citenamefont {Reck}\ \emph {et~al.}(1994)\citenamefont {Reck},
  \citenamefont {Zeilinger}, \citenamefont {Bernstein},\ and\ \citenamefont
  {Bertani}}]{Reck1994}%
  \BibitemOpen
  \bibfield  {author} {\bibinfo {author} {\bibfnamefont {M.}~\bibnamefont
  {Reck}}, \bibinfo {author} {\bibfnamefont {A.}~\bibnamefont {Zeilinger}},
  \bibinfo {author} {\bibfnamefont {H.~J.}\ \bibnamefont {Bernstein}},\ and\
  \bibinfo {author} {\bibfnamefont {P.}~\bibnamefont {Bertani}},\ }\bibfield
  {title} {\bibinfo {title} {Experimental realization of any discrete unitary
  operator},\ }\href {https://doi.org/10.1103/PhysRevLett.73.58} {\bibfield
  {journal} {\bibinfo  {journal} {Phys. Rev. Lett.}\ }\textbf {\bibinfo
  {volume} {73}},\ \bibinfo {pages} {58} (\bibinfo {year} {1994})}\BibitemShut
  {NoStop}%
\bibitem [{\citenamefont {Clements}\ \emph {et~al.}(2016)\citenamefont
  {Clements}, \citenamefont {Humphreys}, \citenamefont {Metcalf}, \citenamefont
  {Kolthammer},\ and\ \citenamefont {Walmsley}}]{Clements2016}%
  \BibitemOpen
  \bibfield  {author} {\bibinfo {author} {\bibfnamefont {W.~R.}\ \bibnamefont
  {Clements}}, \bibinfo {author} {\bibfnamefont {P.~C.}\ \bibnamefont
  {Humphreys}}, \bibinfo {author} {\bibfnamefont {B.~J.}\ \bibnamefont
  {Metcalf}}, \bibinfo {author} {\bibfnamefont {W.~S.}\ \bibnamefont
  {Kolthammer}},\ and\ \bibinfo {author} {\bibfnamefont {I.~A.}\ \bibnamefont
  {Walmsley}},\ }\bibfield  {title} {\bibinfo {title} {Optimal design for
  universal multiport interferometers},\ }\href
  {https://doi.org/10.1364/OPTICA.3.001460} {\bibfield  {journal} {\bibinfo
  {journal} {Optica}\ }\textbf {\bibinfo {volume} {3}},\ \bibinfo {pages}
  {1460} (\bibinfo {year} {2016})}\BibitemShut {NoStop}%
\bibitem [{\citenamefont {van Loock}\ and\ \citenamefont
  {L{\"u}tkenhaus}(2004)}]{vanLoockLutkenhaus2004}%
  \BibitemOpen
  \bibfield  {author} {\bibinfo {author} {\bibfnamefont {P.}~\bibnamefont {van
  Loock}}\ and\ \bibinfo {author} {\bibfnamefont {N.}~\bibnamefont
  {L{\"u}tkenhaus}},\ }\bibfield  {title} {\bibinfo {title} {Simple criteria
  for the implementation of projective measurements with linear optics},\
  }\href {https://doi.org/10.1103/PhysRevA.69.012302} {\bibfield  {journal}
  {\bibinfo  {journal} {Phys. Rev. A}\ }\textbf {\bibinfo {volume} {69}},\
  \bibinfo {pages} {012302} (\bibinfo {year} {2004})}\BibitemShut {NoStop}%
\bibitem [{\citenamefont {Horn}\ and\ \citenamefont
  {Johnson}(2013)}]{HornJohnson2013}%
  \BibitemOpen
  \bibfield  {author} {\bibinfo {author} {\bibfnamefont {R.~A.}\ \bibnamefont
  {Horn}}\ and\ \bibinfo {author} {\bibfnamefont {C.~R.}\ \bibnamefont
  {Johnson}},\ }\href@noop {} {\emph {\bibinfo {title} {Matrix Analysis}}},\
  \bibinfo {edition} {2nd}\ ed.\ (\bibinfo  {publisher} {Cambridge University
  Press},\ \bibinfo {year} {2013})\BibitemShut {NoStop}%
\bibitem [{\citenamefont {Goyal}\ \emph {et~al.}(2014)\citenamefont {Goyal},
  \citenamefont {Boukama-Dzoussi}, \citenamefont {Ghosh}, \citenamefont
  {Roux},\ and\ \citenamefont {Konrad}}]{Goyal2014}%
  \BibitemOpen
  \bibfield  {author} {\bibinfo {author} {\bibfnamefont {S.~K.}\ \bibnamefont
  {Goyal}}, \bibinfo {author} {\bibfnamefont {P.~E.}\ \bibnamefont
  {Boukama-Dzoussi}}, \bibinfo {author} {\bibfnamefont {S.}~\bibnamefont
  {Ghosh}}, \bibinfo {author} {\bibfnamefont {F.~S.}\ \bibnamefont {Roux}},\
  and\ \bibinfo {author} {\bibfnamefont {T.}~\bibnamefont {Konrad}},\
  }\bibfield  {title} {\bibinfo {title} {Qudit-teleportation for photons with
  linear optics},\ }\href {https://doi.org/10.1038/srep04543} {\bibfield
  {journal} {\bibinfo  {journal} {Sci. Rep.}\ }\textbf {\bibinfo {volume}
  {4}},\ \bibinfo {pages} {4543} (\bibinfo {year} {2014})}\BibitemShut
  {NoStop}%
\bibitem [{\citenamefont {Zhang}\ \emph
  {et~al.}(2019{\natexlab{b}})\citenamefont {Zhang}, \citenamefont {Chen},
  \citenamefont {Cui}, \citenamefont {Dowling}, \citenamefont {Ou},\ and\
  \citenamefont {Byrnes}}]{ZhangTeleport2019}%
  \BibitemOpen
  \bibfield  {author} {\bibinfo {author} {\bibfnamefont {C.}~\bibnamefont
  {Zhang}}, \bibinfo {author} {\bibfnamefont {J.~F.}\ \bibnamefont {Chen}},
  \bibinfo {author} {\bibfnamefont {C.}~\bibnamefont {Cui}}, \bibinfo {author}
  {\bibfnamefont {J.~P.}\ \bibnamefont {Dowling}}, \bibinfo {author}
  {\bibfnamefont {Z.~Y.}\ \bibnamefont {Ou}},\ and\ \bibinfo {author}
  {\bibfnamefont {T.}~\bibnamefont {Byrnes}},\ }\bibfield  {title} {\bibinfo
  {title} {Quantum teleportation of photonic qudits using linear optics},\
  }\href {https://doi.org/10.1103/PhysRevA.100.032330} {\bibfield  {journal}
  {\bibinfo  {journal} {Phys. Rev. A}\ }\textbf {\bibinfo {volume} {100}},\
  \bibinfo {pages} {032330} (\bibinfo {year} {2019}{\natexlab{b}})}\BibitemShut
  {NoStop}%
\end{thebibliography}%

\clearpage
\onecolumngrid
\setcounter{secnumdepth}{2}

\setcounter{section}{0}
\setcounter{subsection}{0}
\setcounter{secnumdepth}{2}
\setcounter{equation}{0}
\setcounter{figure}{0}
\setcounter{table}{0}
\renewcommand{\thesection}{S\arabic{section}}
\renewcommand{\thesubsection}{S\arabic{section}.\arabic{subsection}}
\renewcommand{\theequation}{S\arabic{equation}}
\renewcommand{\thefigure}{S\arabic{figure}}
\renewcommand{\thetable}{S\arabic{table}}
\renewcommand{\theHsection}{S.\arabic{section}}
\renewcommand{\theHsubsection}{S.\arabic{section}.\arabic{subsection}}
\renewcommand{\theHequation}{S.\arabic{equation}}
\renewcommand{\theHfigure}{S.\arabic{figure}}
\renewcommand{\theHtable}{S.\arabic{table}}

\begin{center}
{\large\bf Supplemental Material for ``Auxiliary Schmidt Rank as a Resource for Photonic Bell Measurements''}\par\vspace{0.75em}
Pradip Laha and Peter van Loock\\
{\it Institute of Physics, Johannes Gutenberg-Universit\"at Mainz, Staudingerweg 7, 55128 Mainz, Germany}
\end{center}

\section{Measurement model and fine-grained outcomes}
\label{sec:supp_model}

The main text considers a two-photon, same-photon-assisted Bell-label decoding task.  The notation \(R_A,R_B\) may denote either one auxiliary degree of freedom or the tensor product of several auxiliary degrees of freedom carried by the same two photons:
\[
\mathcal{H}_{R_A}=\bigotimes_\kappa \mathcal{H}_{R_A^{(\kappa)}},
\qquad
\mathcal{H}_{R_B}=\bigotimes_\kappa \mathcal{H}_{R_B^{(\kappa)}} .
\]
The resource used in the theorem is the Schmidt rank \(r_\Phi\) of
the fixed auxiliary state across the combined \(R_A|R_B\) partition.
If \(\ket{\Phi}=\bigotimes_\kappa\ket{\Phi^{(\kappa)}}\), then
\[
r_\Phi=\prod_\kappa r_{\Phi^{(\kappa)}} .
\]
Thus, several auxiliary degrees of freedom do not change the theorem; they provide a larger combined auxiliary Hilbert space in which the required Schmidt rank may be encoded.

The input ensemble is
\begin{equation}
  \mathcal{S}^{(d)}_{\Phi} = \left\{ \ket{\Psi^{(d)}_{pq}}_{S_A S_B}\otimes \ket{\Phi}_{R_A R_B}:p,q\in\mathbb{Z}_d \right\},
\label{eq:supp_ensemble}
\end{equation}
where the auxiliary state $\ket{\Phi}$ is fixed (known) and independent of the unknown system Bell label. The measurement consists of one passive linear-optical unitary on all modes, followed by ideal PNR detection and classical postprocessing. Extra modes may be present, but they are initially vacuum. No additional photons are populated.

A deterministic decoder assigns every fine-grained photon-number pattern $\alpha$ either to one Bell label $\ell=(p,q)$ or to an impossible outcome. If $\alpha$ is assigned to $\ell$, then the conditional fine-grained detection probability
\begin{equation}
P(\alpha|\ell')=0
\qquad
\text{for all } \ell'\neq \ell,
\label{eq:supp_finegrained}
\end{equation}
whenever $P(\alpha|\ell)>0$. This is the microscopic form of unambiguous deterministic Bell label readout.

Coarse-graining cannot evade the no-go theorem. If a coarse-grained outcome $s$ is conclusive for label $\ell$, then
\begin{equation}
P(s|\ell') = \sum_{\alpha\in s} P(\alpha|\ell') = 0 \qquad (\ell'\neq \ell).
\label{eq:supp_coarse}
\end{equation}
Since each summand is nonnegative, every fine-grained pattern inside $s$ is itself zero on every non-target label. Thus, it is sufficient and strongest to prove the rank obstruction at the ideal photon-number-resolved level. Threshold and finite-resolution detectors are classical mergings of these fine-grained outcomes.

\section{Two-photon amplitude algebra and the bare-qudit obstruction}
\label{sec:supp_bare}

This section gives the formal version of the bare-qudit obstruction used in the main text.  Let the populated logical input registers be
\[
A=\{a_0^\dagger,\ldots,a_{d-1}^\dagger\},
\qquad
B=\{b_0^\dagger,\ldots,b_{d-1}^\dagger\}.
\]
Any pure bare two-qudit state with one photon in each register can be written as
\begin{equation}
\ket{C}
=
\sum_{m,n=0}^{d-1}
C_{mn}a_m^\dagger b_n^\dagger\ket{\mathrm{vac}},
\qquad
\Tr(C^\dagger C)=1 .
\label{eq:supp_C}
\end{equation}
The Schmidt rank of \(\ket C\) across the \(A|B\) partition is \(\rank C\).

For a fine-grained two-photon output pattern \((\mu,\nu)\), the normalized detection bra is
\begin{equation}
  {}_{c}\!\bra{\mu,\nu} = \frac{\bra{\mathrm{vac}}\,c_\mu c_\nu} {\sqrt{1+\delta_{\mu\nu}}}.
  \label{eq:supp_detection_bra}
\end{equation}
The denominator accounts for the normalization of a two-photon Fock state when both photons occupy the same output mode. 
The passive linear-optical mode transformation expresses each output annihilation operator as a linear combination of all input annihilation operators.  Keeping only the coefficients multiplying the populated logical modes $a_m$ and $b_n$, we write
\begin{equation}
  c_\mu = \sum_{m=0}^{d-1}(\alpha_\mu)_m a_m + \sum_{n=0}^{d-1}(\beta_\mu)_n b_n +\cdots .
\end{equation}
Here $\alpha_\mu$ and $\beta_\mu$ are column vectors collecting the components of output row $\mu$ on the populated $A$ and $B$ registers, respectively.  The ellipsis naturally denotes annihilation operators for all other input modes.  These modes are initially in vacuum, so any term involving them gives zero when evaluated on the bare two-qudit state $\ket{C}$.

Using the bosonic commutation relations, only the terms in which one output annihilator removes the $A$ photon and the other removes the $B$ photon survive.  Hence, the amplitude of this fine-grained detection event on $\ket{C}$ is
\begin{align}
 A_{\mu,\nu}(C) &=  {}_{c}\!\braket{\mu,\nu|C} 
  = \frac{1}{\sqrt{1+\delta_{\mu\nu}}} \sum_{m,n}C_{mn} \bra{\mathrm{vac}}c_\mu c_\nu a_m^\dagger b_n^\dagger \ket{\mathrm{vac}} \nonumber\\
  &= \frac{1}{\sqrt{1+\delta_{\mu\nu}}} \sum_{m,n}C_{mn} \left[ (\alpha_\mu)_m(\beta_\nu)_n + (\alpha_\nu)_m(\beta_\mu)_n \right] \nonumber\\
  &= \frac{1} {\sqrt{1+\delta_{\mu\nu}}} \left(\alpha_\mu^T C\beta_\nu+\alpha_\nu^T C\beta_\mu\right).
 \label{eq:supp_bare_amp}
\end{align}
The two terms are the two indistinguishable alternatives: $\mu$ annihilates the $A$ photon and $\nu$ the $B$ photon, or $\nu$ annihilates the $A$ photon and $\mu$ the $B$ photon. Equivalently, \( A_{\mu,\nu}(C) = \Tr(P_{\mu\nu}^{T}C)\), where the effective detection matrix is
\begin{equation}
P_{\mu\nu} = \frac{\alpha_\mu\beta_\nu^T+\alpha_\nu\beta_\mu^T}
{\sqrt{1+\delta_{\mu\nu}}}.
\label{eq:supp_bareP}
\end{equation}
Thus, \(P_{\mu\nu}\) is a sum of two outer products, and hence $\rank P_{\mu\nu}\le 2$.
For same-mode events, \(P_{\mu\mu}=\sqrt2\,\alpha_\mu\beta_\mu^T\), so the rank is at most one.  This is the two-photon origin of the Schmidt-rank obstruction: a fine-grained two-click effect is generated by a vector of Schmidt rank at most two.

This rank bound already excludes any bare conclusive generalized Bell-label outcome for $d>2$.  If a fine-grained pattern $(\mu,\nu)$ were unambiguous for label $(p,q)$, then the functional $C\mapsto \Tr(P_{\mu\nu}^{T}C)$ would have to vanish on all Bell matrices except $C_{pq}$.  Since the Bell matrices form a complete Hilbert-Schmidt basis, this forces $P_{\mu\nu}$ to be proportional to the dual Bell matrix $C_{pq}^{*}$, up to normalization.  But $C_{pq}$, and hence $C_{pq}^{*}$, has rank $d$, whereas $\rank P_{\mu\nu}\le2$.  Thus, no microscopic bare two-click pattern is conclusive for $d>2$.  By the coarse-graining argument of Sec.~\ref{sec:supp_model}, the strict bare two-photon/vacuum-mode model therefore admits no conclusive generalized Bell-label outcome for $d>2$.  This is the two-photon specialization of Calsamiglia's linear-optical Schmidt-number obstruction~\cite{Calsamiglia2002}, with the bare-qudit implication emphasized by Du\v{s}ek~\cite{Dusek2001}.

\section{Contraction space and the single-outcome threshold}
\label{sec:supp_contraction}

We assume that each photon carries a system degree of freedom and one or more auxiliary degrees of freedom.  We write the combined one-photon spaces as
\[
\mathcal{H}_A=\mathcal{H}_{S_A}\otimes\mathcal{H}_{R_A},
\qquad
\mathcal{H}_B=\mathcal{H}_{S_B}\otimes\mathcal{H}_{R_B}.
\]
The fixed auxiliary state on the combined auxiliary spaces is
\begin{equation}
  \ket{\Phi}_{R_A R_B} = \sum_{a,b}\Phi_{ab}\ket a_{R_A}\ket b_{R_B},
  \qquad r_\Phi=\rank\Phi .
  \label{eq:supp_phi}
\end{equation}
Here \(r_\Phi\) is the Schmidt rank of the auxiliary state across \(R_A|R_B\); the state is entangled iff \(r_\Phi>1\).

For clarity, write the system input as
\[
  \ket C_{S_A S_B} =  \sum_{i,j}C_{ij}\ket i_{S_A}\ket j_{S_B}.
\]
On the enlarged bipartition
$(S_A\otimes R_A),|,(S_B\otimes R_B)$, 
we use the combined labels \(\ket{i,a}_A=\ket i_{S_A}\ket a_{R_A}\) and \(\ket{j,b}_B=\ket j_{S_B}\ket b_{R_B}\).  A microscopic two-click detection vector on this enlarged space has coefficient matrix \(P\),
\[
  \ket P = \sum_{i,j,a,b} P_{(i,a),(j,b)}
\ket{i,a}_A\ket{j,b}_B .
\]
Its amplitude on the known auxiliary input is
\begin{align}
\braket{P|C\otimes\Phi}
&=
\sum_{i,j,a,b}
P_{(i,a),(j,b)}^*\,C_{ij}\Phi_{ab}
=
\sum_{i,j}
\left[
\sum_{a,b}
P_{(i,a),(j,b)}\Phi_{ab}^*
\right]^*
C_{ij}.
\end{align}
This motivates the contracted system-level detection matrix
\begin{equation}
[\Gamma_\Phi(P)]_{ij} = \sum_{a,b}
P_{(i,a),(j,b)}\Phi_{ab}^{*},
\label{eq:supp_Gamma}
\end{equation}
for which
\begin{equation}
  \braket{P|C\otimes\Phi} = \Tr\!\left[\Gamma^\dagger_\Phi(P) \, C\right].
  \label{eq:supp_Gamma_amp}
\end{equation}
Thus, \(\Gamma_\Phi(P)\) is the Bell-label functional seen by the system state $\ket{C}$ after the known auxiliary state has been inserted.  The complex conjugation on \(\Phi\) in Eq.~\eqref{eq:supp_Gamma} is a convention chosen so that
\[
\braket{P|C\otimes\Phi} =  \sum_{i,j}[\Gamma_\Phi(P)]_{ij}^*C_{ij} = \Tr[\Gamma_\Phi^\dagger(P) C],
\]
the standard Hilbert-Schmidt inner product between the matrices \(\Gamma_\Phi(P)\) and \(C\).

A microscopic two-click detection vector on the enlarged bipartition \((S_A R_A)|(S_B R_B)\) still has Schmidt rank at most two.  Equivalently, its coefficient matrix \(P\), whose row index is \((i,a\)) and column index is \((j,b)\), has rank at most two and can be written as a sum of two outer products, given by
\begin{equation}
  P=u_1v_1^T+u_2v_2^T .
\label{eq:supp_Pdecomp}
\end{equation}
Here \(u_t\) and \(v_t\) are vectors in the enlarged one-photon spaces on the two sides \(A\) and \(B\), respectively.  After restricting the auxiliary labels to the Schmidt support of \(\ket{\Phi}\), these vectors have length \(d r_\Phi\).  We then reshape them into \(d\times r_\Phi\) matrices,
\[
(U_t)_{ia}=(u_t)_{(i,a)},\qquad (V_t)_{jb}=(v_t)_{(j,b)} ,
\]
so we have moved from the vector representation of the enlarged two-click effect to matrices indexed by system and auxiliary labels.  Substituting
\[
P_{(i,a),(j,b)} =  \sum_{t=1}^2 (U_t)_{ia}(V_t)_{jb}
\]
into Eq.~\eqref{eq:supp_Gamma} gives
\begin{equation}
\Gamma_\Phi(P) =  \sum_{t=1}^{2} U_t\Phi^{*}V_t^T .
\label{eq:supp_contract}
\end{equation}
Thus, each rank-one enlarged term can contribute system rank at most \(r_\Phi\), so
\[
\rank\Gamma_\Phi(P)\le \min(d,2r_\Phi).
\]
Conversely, every $d\times d$ matrix $Q$ with $\mathrm{rank}\,Q\leqslant \min(d,2r_\Phi)$
can be represented in the form~\eqref{eq:supp_contract}. To see this, split a singular-value decomposition of $Q$ into two matrices $Q_1,Q_2$, each of rank at most $r_\Phi$, and factor each through the rank-$r_\Phi$ auxiliary matrix:
\[
Q_t=U_t\Phi^*V_t^T \qquad (t=1,2).
\]
Vectorizing $U_t,V_t$ gives a Schmidt-rank-two enlarged matrix $P=u_1v_1^T+u_2v_2^T$ with $\Gamma_\Phi(P)=Q$. Therefore, the rank-relaxed contraction space is
\begin{equation}
\mathcal{K}^{(2)}_\Phi = \left\{Q\in\mathbb{C}^{d\times d}: \mathrm{rank}\,Q\leqslant \min(d,2r_\Phi) \right\}.
\label{eq:supp_rank_variety}
\end{equation}
Equation~\eqref{eq:supp_rank_variety} characterizes the algebraic contraction space only. It does not assert that arbitrary elements of this space can be realized as mutually compatible click patterns of a single passive interferometer.

A microscopic outcome conclusive for label $(p,q)$ must be orthogonal to all Bell matrices except $C_{pq}$. Since the $C_{pq}$ form a complete orthonormal matrix basis, this requires
\begin{equation}
\Gamma_\Phi(P)\propto C_{pq}.
\label{eq:supp_conclusive_proportional}
\end{equation}
Because $C_{pq}$ has rank $d$, a single conclusive Bell-label functional exists in this rank relaxation if and only if
\begin{equation}
2r_\Phi\geqslant d, 
\qquad \text{or equivalently} \qquad
r_\Phi\geqslant \left\lceil\frac d2\right\rceil .
\label{eq:supp_single_threshold}
\end{equation}
This is a single-outcome threshold. It does not guarantee that all such contractions can be realized simultaneously by one passive interferometer as a deterministic photon-counting analyzer.

\section{Deterministic full-rank theorem}
\label{sec:supp_deterministic}

Put the auxiliary state in Schmidt form on its support,
\begin{equation}
\ket{\Phi}_{R_A R_B} = \sum_{a=1}^{r}\lambda_a \ket a_{R_A}\ket a_{R_B}, \qquad r=r_\Phi,
\label{eq:supp_schmidt}
\end{equation}
with $\lambda_a>0$, and define
\[
\Lambda=\mathrm{diag}(\lambda_1,\ldots,\lambda_r).
\]
For output mode $\mu$, let $X_\mu,Y_\mu\in\mathbb{C}^{d\times r}$ be the restrictions of the interferometer row to the populated $A$ and $B$ input registers, restricted to the auxiliary Schmidt support. The system-level contraction for the fine-grained output pattern $(\mu,\nu)$ is
\begin{equation}
Q_{\mu\nu} = \frac{ X_\mu\Lambda Y_\nu^T+ X_\nu\Lambda Y_\mu^T } {\sqrt{1+\delta_{\mu\nu}}}.
\label{eq:supp_Q}
\end{equation}

In a deterministic decoder, every nonzero fine-grained outcome must be assigned to a unique Bell label. Hence, every nonzero $Q_{\mu\nu}$ must be proportional to one of the full-rank Bell matrices $C_{pq}$. Assume, for contradiction, that $r<d$.

For a same-mode event,
\begin{equation}
Q_{\mu\mu} = \sqrt{2}\,X_\mu\Lambda Y_\mu^T.
\label{eq:supp_same}
\end{equation}
This has rank at most $r<d$, so it cannot be proportional to a Bell matrix. Determinism therefore forces
\begin{equation}
X_\mu\Lambda Y_\mu^T=0 \qquad \text{for every } \mu .
\label{eq:supp_same_zero}
\end{equation}
Let
\[
a_\mu=\mathrm{rank}\,X_\mu,
\qquad
b_\mu=\mathrm{rank}\,Y_\mu .
\]
Since \(\Lambda\) is invertible on the auxiliary Schmidt support, \(\operatorname{rank}(X_\mu\Lambda)=\operatorname{rank}X_\mu\). Sylvester's inequality~\cite{HornJohnson2013} gives
\begin{equation}
0 = \mathrm{rank}(X_\mu\Lambda Y_\mu^T) \ge a_\mu+b_\mu-r.
\label{eq:supp_sylvester}
\end{equation}
Thus
\begin{equation}
a_\mu+b_\mu\leqslant r \qquad \text{for every } \mu .
\label{eq:supp_rank_split}
\end{equation}

For an off-diagonal event $\mu\neq\nu$,
\begin{align}
\mathrm{rank}\,Q_{\mu\nu} &\le \mathrm{rank}(X_\mu\Lambda Y_\nu^T) + \mathrm{rank}(X_\nu\Lambda Y_\mu^T) \nonumber\\
&\le \min(a_\mu,b_\nu)+\min(a_\nu,b_\mu) \nonumber\\
&\le \frac{a_\mu+b_\nu+a_\nu+b_\mu}{2} \nonumber\\
&= \frac{(a_\mu+b_\mu)+(a_\nu+b_\nu)}{2} \leqslant r < d .
\label{eq:supp_offdiag}
\end{align}
Therefore, no off-diagonal contraction can be proportional to a full-rank Bell matrix either. Same-mode events vanish, and off-diagonal events are rank deficient. Hence, no microscopic photon-counting outcome can be conclusive, contradicting deterministic decoding of all $d^2$ labels. Thus, $r_\Phi\geqslant d$
is necessary for deterministic same-photon-assisted Bell-label decoding in the stated two-photon passive model.

\section{Tightness construction and boundary examples}
\label{sec:supp_tightness}

\subsection{Saturating construction}
\label{sec:supp_saturating_construction}

The lower bound is tight in the following existential sense: no auxiliary state with \(r_\Phi<d\) can work, while at least one state with \(r_\Phi=d\) does.  The \(d\) Schmidt labels may be physical labels of one auxiliary degree of freedom or composite labels inside several combined auxiliary degrees of freedom.  Take the maximally entangled auxiliary state
\begin{equation}
  \ket{\Phi_d}_{R_A R_B} = \frac{1}{\sqrt d}
  \sum_{a=0}^{d-1}
  \ket a_{R_A}\ket a_{R_B},
\label{eq:supp_phid}
\end{equation}
and define the single-photon Bell basis between the system and auxiliary degrees of freedom by
\begin{equation}
  \ket{\chi_{mn}}_{SR} = 
  \frac{1}{\sqrt d}
  \sum_{t=0}^{d-1}
  \omega^{nt}
  \ket t_S\ket{t+m}_R ,
  \qquad m,n\in\mathbb Z_d .
\label{eq:supp_chi}
\end{equation}
The inverse Fourier relation is
\begin{equation}
  \ket{s}_S\ket{s+m}_R = \frac{1}{\sqrt d} \sum_{n=0}^{d-1} \omega^{-ns}\ket{\chi_{mn}}_{SR}.
  \label{eq:supp_chi_inverse}
\end{equation}

With the Bell-state convention $\ket{\Psi^{(d)}_{pq}}_{S_A S_B} = \frac{1}{\sqrt d} \sum_{j=0}^{d-1} \omega^{qj} \ket{j+p}_{S_A}\ket j_{S_B}$,
the joint system--auxiliary input is
\begin{equation}
  \ket{\Psi^{(d)}_{pq}}_{S_A S_B}\ket{\Phi_d}_{R_A R_B} =
  \frac1d \sum_{j,a} \omega^{qj}
  \ket{j+p}_{S_A}\ket j_{S_B} \ket a_{R_A}\ket a_{R_B}.
  \label{eq:supp_input_expand}
\end{equation}
For each term, introduce the local \(S_A R_A\) Bell-basis shift
\[
m=a-j-p .
\]
Using Eq.~\eqref{eq:supp_chi_inverse} on the two local pairs we get
\begin{align}
\ket{j+p}_{S_A}\ket a_{R_A} &= \ket{j+p}_{S_A}\ket{j+p+m}_{R_A} = 
\frac1{\sqrt d} \sum_n \omega^{-n(j+p)} \ket{\chi_{mn}}_{S_A R_A},\\
\ket j_{S_B}\ket a_{R_B}
&= \ket{j}_{S_A}\ket{j+m+p}_{R_A} = \frac1{\sqrt d} \sum_{n'} \omega^{-n'j} \ket{\chi_{m+p,n'}}_{S_B R_B}.
\end{align}
Substituting these two local relations into Eq.~\eqref{eq:supp_input_expand} yields
\begin{align}
\ket{\Psi^{(d)}_{pq}}\ket{\Phi_d}
&=
\frac{1}{d^2}
\sum_{j,m,n,n'}
\omega^{qj-n(j+p)-n'j}
\ket{\chi_{mn}}_{S_A R_A}
\ket{\chi_{m+p,n'}}_{S_B R_B}
\nonumber\\
&=
\frac{1}{d^2}
\sum_{m,n,n'}
\omega^{-np}
\left[
\sum_j \omega^{j(q-n-n')}
\right]
\ket{\chi_{mn}}_{S_A R_A}
\ket{\chi_{m+p,n'}}_{S_B R_B}.
\end{align}
The bracketed Fourier sum is \(d\,\delta_{n',q-n}\).  Therefore
\begin{equation}
\ket{\Psi^{(d)}_{pq}}_{S_A S_B}
\ket{\Phi_d}_{R_A R_B} = \frac1d
\sum_{m,n\in\mathbb Z_d}
\omega^{-np}
\ket{\chi_{mn}}_{S_A R_A}
\ket{\chi_{m+p,q-n}}_{S_B R_B}.
\label{eq:supp_swap}
\end{equation}

Thus, local Bell-basis outcomes \((m,n)\) on photon \(A\) and \((m',n')\) on photon \(B\) determine the original system Bell label by
\begin{equation}
p=m'-m,
\qquad
q=n+n'
\pmod d .
\label{eq:supp_recovery}
\end{equation}
Because \({\ket{\chi_{mn}}}\) is an orthonormal basis of one photon's \(S\otimes R\) mode space, an ideal passive single-particle unitary can map this basis to distinct output modes~\cite{Reck1994,Clements2016}.  Applying this basis change separately in the \(A\) and \(B\) spatial registers, followed by PNR detection, decodes all \(d^2\) labels deterministically.  

\subsection{Boundary cases}
\label{sec:supp_boundary_cases}

For qutrits, \(d=3\), the gap between the single-outcome and deterministic thresholds is already visible.  The single-outcome condition permits \(r_\Phi=2\), since \(2r_\Phi=4\ge3\).  Indeed, for a target qutrit Bell matrix,
\begin{equation}
C_{pq}
=
\frac{1}{\sqrt3}
\sum_{j=0}^{2}
\omega^{qj}
\ket{j+p}\bra j =
\frac{1}{\sqrt3}
\left(
\ket{p}\bra 0 + \omega^{q}
\ket{p+1}\bra 1 +\omega^{2q}
\ket{p+2}\bra 2
\right),
\label{eq:supp_qutrit_C}
\end{equation}
one can split \(C_{pq}=Q^{(1)}_{pq}+Q^{(2)}_{pq}\), where
\begin{align}
Q^{(1)}_{pq}
=
\frac{1}{\sqrt3}
\left(
\ket p\bra0+\omega^q\ket{p+1}\bra1
\right),
\qquad 
Q^{(2)}_{pq}
&=
\frac{\omega^{2q}}{\sqrt3}
\ket{p+2}\bra2 .
\label{eq:supp_qutrit_split2}
\end{align}
Thus, \(Q^{(1)}_{pq}\) and \(Q^{(2)}_{pq}\) have ranks two and one, respectively, and individual conclusive contractions exist for all nine qutrit Bell labels. Deterministic decoding nevertheless fails for \(r_\Phi=2\): same-mode contractions force the rank split of Sec.~\ref{sec:supp_deterministic}, and then every off-diagonal contraction has rank at most two.  No microscopic outcome can carry a rank-three Bell matrix. An explicit example follows next.

\subsubsection{An explicit \(d=3,\ r_\Phi=2\) single-outcome witness}
\label{sec:supp_qutrit_witness}

The preceding rank split can be realized explicitly for a single conclusive outcome.  Take the target label \((p,q)=(0,0)\) and the rank-two auxiliary state
\begin{equation}
\ket{\Phi_2}_{R_A R_B}
=
\frac{1}{\sqrt2}
\left(
\ket{0}_{R_A}\ket{0}_{R_B}
+
\ket{1}_{R_A}\ket{1}_{R_B}
\right),
\qquad
\Phi=\frac{1}{\sqrt2}I_2 .
\label{eq:supp_phi2}
\end{equation}
Then $C_{00} =
\frac{1}{\sqrt3} I_3 = Q^{(1)}+Q^{(2)}$,
where
\begin{equation}
Q^{(1)}
=
\frac{1}{\sqrt3}
\begin{pmatrix}
1&0&0\\
0&1&0\\
0&0&0
\end{pmatrix},
\qquad
Q^{(2)}
=
\frac{1}{\sqrt3}
\begin{pmatrix}
0&0&0\\
0&0&0\\
0&0&1
\end{pmatrix}.
\label{eq:supp_Q12}
\end{equation}
Thus, \(\rank Q^{(1)}=2\) and \(\rank Q^{(2)}=1\).  With \(\eta=(2/3)^{1/4}\), one possible factorization through the rank-two auxiliary support is
\begin{equation}
U_1=V_1
=
\eta
\begin{pmatrix}
1&0\\
0&1\\
0&0
\end{pmatrix},
\qquad
U_2=V_2
=
\eta
\begin{pmatrix}
0&0\\
0&0\\
1&0
\end{pmatrix}.
\label{eq:supp_UV_witness}
\end{equation}
Since \(\Phi^*=I_2/\sqrt2\),
\begin{equation}
U_1\Phi^*V_1^T=Q^{(1)},
\qquad
U_2\Phi^*V_2^T=Q^{(2)}.
\end{equation}
Hence, the enlarged Schmidt-rank-two witness \(P=u_1v_1^T+u_2v_2^T\) satisfies
\begin{equation}
\Gamma_\Phi(P)
=
U_1\Phi^*V_1^T+U_2\Phi^*V_2^T
=
C_{00}.
\label{eq:supp_witness_gamma}
\end{equation}
Consequently,
\begin{equation}
\langle P|C_{pq}\otimes\Phi_2\rangle
=
\Tr(C_{00}^\dagger C_{pq})
=
\delta_{p0}\delta_{q0}.
\label{eq:supp_witness_conclusive}
\end{equation}
This microscopic outcome is therefore unambiguous for \(\Psi^{(3)}_{00}\): if it occurs, the input Bell label is known to be \((0,0)\).

The same witness can be embedded directly into normalized output modes. Label the enlarged input creation operators by \(A_{sr}^\dagger\) and \(B_{sr}^\dagger\), where \(s=0,1,2\) is the system index and \(r=0,1\) is the auxiliary index.  Define two output modes
\begin{align}
c_\mu^\dagger = \frac{1}{\sqrt3}
\left(
A_{00}^\dagger+A_{11}^\dagger+B_{20}^\dagger
\right),\qquad
c_\nu^\dagger =
\frac{1}{\sqrt3}
\left(
A_{20}^\dagger+B_{00}^\dagger+B_{11}^\dagger
\right).
\end{align}
These two rows are normalized and mutually orthogonal because their input supports are disjoint, and they can be completed to a passive interferometer.  For the two-click pattern \((\mu,\nu)\), the row restrictions are
\begin{equation}
X_\mu=
\frac{1}{\sqrt3}
\begin{pmatrix}
1&0\\
0&1\\
0&0
\end{pmatrix},
\quad
Y_\mu=
\frac{1}{\sqrt3}
\begin{pmatrix}
0&0\\
0&0\\
1&0
\end{pmatrix},
\end{equation}
and
\begin{equation}
X_\nu=
\frac{1}{\sqrt3}
\begin{pmatrix}
0&0\\
0&0\\
1&0
\end{pmatrix},
\quad
Y_\nu=
\frac{1}{\sqrt3}
\begin{pmatrix}
1&0\\
0&1\\
0&0
\end{pmatrix}.
\end{equation}
Therefore,
\begin{equation}
Q_{\mu\nu}
=
X_\mu\Phi^*Y_\nu^T
+
X_\nu\Phi^*Y_\mu^T
=
\frac{1}{3\sqrt2}I_3
=
\frac{1}{\sqrt6}\,C_{00}.
\label{eq:supp_Qmunu_witness}
\end{equation}
Thus, the normalized detector pattern with one photon in output mode \(\mu\) and one photon in output mode \(\nu\) has nonzero amplitude on \(\Psi^{(3)}_{00}\) and zero amplitude on all other qutrit Bell labels:
\begin{equation}
{}_{c}\!\langle \mu,\nu|
\Psi^{(3)}_{pq}\otimes\Phi_2\rangle
=
\frac{1}{\sqrt6}\,\delta_{p0}\delta_{q0}.
\label{eq:supp_click_probability}
\end{equation}
The two monitored modes are balanced three-mode superpositions and may be implemented as tritter outputs, or decomposed into ordinary two-mode beam splitters and phase shifters.  This construction realizes one conclusive microscopic pattern only.  It cannot be completed into a deterministic qutrit analyzer with \(r_\Phi=2\), as shown by the rank-split theorem.


\subsection{Explicit qutrit sorter and distinguishability check}
\label{sec:supp_qutrit_sorter}

For \(d=3\), the saturating construction can be written as an explicit local mode sorter.  The word ``local'' means that the unitary acts on the \(S\otimes R\) degrees of freedom of one photon only; the same unitary is then applied independently to photons \(A\) and \(B\).  Label one photon's computational modes by
\[
\ket{s,r}\equiv \ket{s}_{S}\ket{r}_{R},
\qquad
s,r\in\mathbb Z_3 .
\]
The goal is to sort the single-photon Bell basis $\ket{\chi_{mn}} = \frac{1}{\sqrt3} \sum_{s=0}^{2} \omega^{ns}\ket{s,s+m}$
into distinct output modes labelled by \((n,m)\).  This can be done in two elementary steps.  First, apply the permutation
\begin{equation}
\Delta:\ket{s,r}\mapsto \ket{s,r-s},
\label{eq:supp_delta}
\end{equation}
which stores the relative label \(m=r-s\) in the second register.  Thus,
\[
\Delta\ket{\chi_{mn}}
=\left(
\frac{1}{\sqrt3}
\sum_{s=0}^{2}
\omega^{ns}\ket{s}
\right)
\ket m .
\]
Second, apply the inverse qutrit Fourier transform \(F_3^\dagger\) to the first register, with
\[
\bra{n}F_3^\dagger\ket{s}
=\frac{1}{\sqrt3}\omega^{-ns}.
\]
This maps the Fourier superposition above to \(\ket n\).  Hence, the local sorter is
\begin{equation}
U_{\rm loc}
=(F_3^\dagger\otimes I_3)\Delta,
\qquad
\bra{n,m}U_{\rm loc}\ket{s,r}
=\frac{1}{\sqrt3}\omega^{-ns}\delta_{m,r-s},
\label{eq:supp_uloc}
\end{equation}
and it satisfies
\begin{equation}
  U_{\rm loc}\ket{\chi_{mn}}=\ket{n,m}.
\end{equation}
Applying this sorter to both photons gives local outcomes \((n_A,m_A)\) and \((n_B,m_B)\).  The recovery rule from the swapping identity is then
\begin{equation}
p=m_B-m_A,
\qquad
q=n_A+n_B
\pmod3 .
\label{eq:supp_qutrit_rule}
\end{equation}
Thus, the \(d=3,\ r_\Phi=3\) construction is a concrete passive single-particle sorting of the local \(S\otimes R\) Bell basis, followed by PNR detection and classical relabelling.

We now verify that this ideal qutrit construction satisfies the
linear-optical exact-distinguishability criteria of van Loock and
L\"utkenhaus \cite{vanLoockLutkenhaus2004}.  Let
\begin{equation}
\ket{\widetilde{\Psi}_{pq}}
=
\ket{\Psi^{(3)}_{pq}}_{S_A S_B}
\ket{\Phi_3}_{R_A R_B},
\qquad
p,q\in\mathbb{Z}_3 .
\end{equation}
The local \(S\otimes R\) sorter maps
\(\ket{\chi_{mn}}_{S_A R_A}\) and
\(\ket{\chi_{mn}}_{S_B R_B}\) to distinct one-photon output modes,
denoted \(f^\dagger_{A;n,m}\) and \(f^\dagger_{B;n,m}\).  Using
Eq.~\eqref{eq:supp_swap}, the output state is
\begin{equation}
\ket{\widetilde{\Psi}^{\rm out}_{pq}}
=
\frac{1}{3}
\sum_{m,n\in\mathbb{Z}_3}
\omega^{-np}
f^\dagger_{A;n,m}
f^\dagger_{B;q-n,m+p}
\ket{\mathrm{vac}} .
\label{eq:supp_vLL_output}
\end{equation}
Thus, each term corresponds to the fine-grained two-photon Fock pattern
\begin{equation}
\mathsf{F}^{pq}_{m,n}
=
(A:n,m;\,B:q-n,m+p).
\label{eq:supp_vLL_pattern}
\end{equation}

The Fock supports for distinct Bell labels are disjoint.  If
\[
\mathsf{F}^{pq}_{m,n}
=
\mathsf{F}^{p'q'}_{m',n'},
\]
then equality of the occupied \(A\)-mode gives \(n=n'\) and \(m=m'\). Equality of the occupied \(B\)-mode then gives
\[
q-n=q'-n',
\qquad
m+p=m'+p',
\]
and hence \(p=p'\) and \(q=q'\).  Therefore two distinct labels \((p,q)\neq(p',q')\) share no photon-number-resolved output pattern.

Let \(\Delta_{\rm Fock}\) denote complete dephasing in the output Fock basis, i.e., the classical information retained by ideal PNR detection. From Eq.~\eqref{eq:supp_vLL_output},
\begin{equation}
\Delta_{\rm Fock}
\!\left(
\ket{\widetilde{\Psi}^{\rm out}_{pq}}\bra{\widetilde{\Psi}^{\rm out}_{pq}}
\right)
=
\frac{1}{9}
\sum_{m,n}
\ket{\mathsf{F}^{pq}_{m,n}}
\bra{\mathsf{F}^{pq}_{m,n}} .
\end{equation}
Because the supports are disjoint,
\begin{equation}
\Tr\!\left[
\Delta_{\rm Fock}(\ket{\widetilde{\Psi}^{\rm out}_{pq}}\bra{\widetilde{\Psi}^{\rm out}_{pq}})
\,
\Delta_{\rm Fock}(\ket{\widetilde{\Psi}^{\rm out}_{p'q'}}\bra{\widetilde{\Psi}^{\rm out}_{p'q'}})
\right]
=
0
\quad
\bigl((p,q)\neq(p',q')\bigr).
\label{eq:supp_vLL_dephased}
\end{equation}
Equivalently, all normally ordered photon-counting moment conditions are satisfied.  For every output mode \(j\),
\begin{equation}
\bra{\widetilde{\Psi}^{\rm out}_{pq}}
\hat n_j
\ket{\widetilde{\Psi}^{\rm out}_{p'q'}}
=
0
\quad
\bigl((p,q)\neq(p',q')\bigr),
\end{equation}
and for every pair of output modes \(j,k\),
\begin{equation}
\bra{\widetilde{\Psi}^{\rm out}_{pq}}
c_j^\dagger c_k^\dagger c_k c_j
\ket{\widetilde{\Psi}^{\rm out}_{p'q'}}
=
0
\quad
\bigl((p,q)\neq(p',q')\bigr).
\end{equation}
These operators are diagonal in the output Fock basis, and the different Bell labels have disjoint Fock support.  Since the code contains exactly two photons, this exhausts the nontrivial photon-number-resolved conditions.

The same conclusion is visible conditionally.  If the first detected photon is in mode \(f^\dagger_{A;n,m}\), then for input label \((p,q)\) the remaining photon is in \(f^\dagger_{B;q-n,m+p}\).  For fixed \((n,m)\), the map
\[
(p,q)\mapsto(q-n,m+p)
\]
is one-to-one over \(\mathbb{Z}_3^2\).  Thus, the conditional one-photon states for different Bell labels are orthogonal.  The same argument holds if the first detected photon is in the \(B\) block. Therefore, the ideal qutrit construction satisfies the van Loock--L\"utkenhaus exact distinguishability criteria, and PNR detection followed by the classical rule in Eq.~\eqref{eq:supp_qutrit_rule} implements deterministic Bell-label decoding on the assisted nine-state code space.

\section{Related tasks, resource audit, and open points}
\label{sec:supp_scope}

This theorem applies only to deterministic full-label decoding in the following resource model: exactly two populated photons, a fixed same-photon auxiliary state, static passive linear optics, arbitrary vacuum modes, PNR detection, and classical postprocessing.  Changing the populated Fock sector, the operation class, or the success criterion produces a different resource class.
A grouped analyzer identifies a subset \(G_s\subseteq \mathbb{Z}_d\times\mathbb{Z}_d\) rather than a unique Bell label.  At the microscopic level this requires
\begin{equation}
\Gamma_\Phi(P)
\in
\mathrm{span}\{C_\ell:\ell\in G_s\},
\label{eq:supp_grouped_condition}
\end{equation}
rather than \(\Gamma_\Phi(P)\propto C_\ell\) for a single label.
Thus, grouped schemes may provide useful partial information without implementing deterministic complete Bell-label decoding.  If labels are uniformly distributed and the grouping is deterministic, a natural task-specific information measure is
\begin{equation}
I(\ell;s)
=
\log(d^2)
-
\frac{1}{d^2}
\sum_s |G_s|\log |G_s|,
\label{eq:supp_grouped_info}
\end{equation}
with the logarithm base chosen according to the desired units.  This should not be reported as complete \(d^2\)-label BSM success probability.

The resource classification relevant to the work is:
\begin{itemize}
\item \emph{Bare two-qudit analyzers:} two populated photons and vacuum auxiliary modes.  For \(d>2\), no conclusive generalized Bell-label outcome exists in the strict model~\cite{Calsamiglia2002,Dusek2001}.

\item \emph{Same-photon-assisted analyzers:} two populated photons and a fixed auxiliary state \(\ket{\Phi}\) carried by the same photons, where \(r_\Phi\) is the Schmidt rank across the combined auxiliary partition.  Deterministic full-label decoding requires \(r_\Phi\ge d\), and \(\ket{\Phi_d}\) saturates the bound in ideal modes.  Existing high-dimensional auxiliary-entanglement constructions use this full-rank auxiliary resource~\cite{Zhang2019}.

\item \emph{Hyperentanglement-assisted qubit analyzers:} assisted polarization-label readout using an auxiliary degree of freedom of the same photon pair~\cite{Kwiat1998,Walborn2003,Schuck2006,Barbieri2007,Wei2007}. These are consistent with the theorem because the assisted qubit case has \(r_\Phi=d=2\).  They are not complete measurements of an unknown enlarged \(S\otimes R\) Bell basis.

\item \emph{Ancillary-photon analyzers:} schemes that populate additional photonic modes before the interferometer~\cite{Grice2011,Ewert2014,Olivo2018,Bayerbach2023,Hauser2025,Baghdasar2025}. These are not constrained by the same two-photon rank theorem because the populated Fock sector has changed.

\item \emph{Squeezed, nonlinear, or otherwise active analyzers:} schemes that modify the measurement algebra through predetection squeezing, nonlinear interactions, or related active operations~\cite{Zaidi2013,Kilmer2019,Bianchi2025,Bianchi2025_PRA}. These are outside the static passive two-photon model.

\item \emph{Logical Bell measurements and fusion operations:} encoded-readout, graph-building, or fusion tasks whose success condition is not deterministic physical \(d^2\)-label readout~\cite{Hilaire2023,Schmidt2024Fusion,Ustun2025,Yamazaki2025,Reiss2026LogicalBM}.

\item \emph{Grouped high-dimensional analyzers:} measurements that identify subsets of Bell labels or partial label information rather than a unique one of all \(d^2\) generalized Bell labels~\cite{Zeng2025}.  Such schemes can be useful for dense-coding or network tasks, but they are not counterexamples to a theorem about complete deterministic Bell-label decoding.
\end{itemize}

Two open points are left by the rank theorem.  First, the sufficiency direction is existential: the maximally entangled auxiliary state \(\ket{\Phi_d}\) works in the ideal mode model, but it remains open whether every full-rank nonmaximally entangled auxiliary state can support deterministic decoding.  Second, implementing the ideal local \(S\otimes R\) Bell-basis unitary is a platform-level synthesis problem; here \(R\) may itself be a combined auxiliary mode space. Path, time-bin, frequency, orbital-angular-momentum, and hybrid encodings may realize the required one-particle unitary with different physical overheads, losses, and mode-matching constraints.  These implementation costs are separate from the auxiliary-Schmidt-rank resource certified in the main text.

\section{Protocol-level consequences}
\label{sec:supp_protocol}

This section records how the rank threshold enters standard BM-based protocols.  We do not attempt to optimize probabilistic or grouped analyzers.  Instead, we isolate the consequence of the theorem for protocols whose ideal operation requires complete readout of one of all \(d^2\) physical Bell labels.

Let \(p_{\rm BM}\) denote the success probability of a complete Bell-label measurement subroutine in a fixed optical resource model, conditioned on the two photons arriving at the analyzer and on the auxiliary state having been prepared.  Here ``success'' means that the outcome identifies a unique label \((p,q)\in\mathbb{Z}_d^2\). In teleportation, this label tells the receiver which inverse generalized Pauli (Weyl) operator, built from the shift \(X\) and phase \(Z\), must be applied to recover the input state. 
In the same-photon assisted passive two-photon model used in the main text,
\begin{equation}
r_\Phi<d
\quad\Longrightarrow\quad
p_{\rm BM}<1
\end{equation}
for deterministic full-label decoding, while the maximally entangled rank-\(d\) auxiliary state gives \(p_{\rm BM}=1\) in the ideal mode model.  In the strict bare two-photon/vacuum-mode model, \(p_{\rm BM}=0\) for complete unambiguous generalized Bell-label readout when \(d>2\)~\cite{Calsamiglia2002}, because no individual generalized Bell label has a conclusive microscopic outcome.

\subsection{Teleportation}

In the standard \(d\)-dimensional teleportation protocol, an unknown input qudit \(\ket{\psi}_{T}\) and Alice's half \(A\) of a shared pair \(\ket{\Phi_d}_{AB}\) are measured jointly in the Bell basis; the other half \(B\) is held by the receiver~\cite{Bennett1993,Bouwmeester1997}. With our choice of Bell-state convention, the teleportation identity takes the form
\begin{equation}
  \ket{\psi}_T\ket{\Phi_d}_{AB} = \frac{1}{d} \sum_{p,q\in\mathbb{Z}_d} \omega^{pq} \ket{\Psi^{(d)}_{pq}}_{TA} X^{-p}Z^{-q}\ket{\psi}_B .
 \label{eq:supp_tel_identity}
\end{equation}
Thus, after outcome \((p,q)\), the receiver applies \(Z^qX^p\) to recover \(\ket{\psi}\), up to an irrelevant global phase.
If the Bell-label readout succeeds with probability \(p_{\rm BM}\), then the ideal heralded teleportation success probability, conditioned on state preparation and photon arrival, is
\begin{equation}
p_{\rm tel}=p_{\rm BM}.
\label{eq:supp_tel_success}
\end{equation}
Additional source, transmission, coupling, and detector efficiencies multiply this factor but do not change the BM resource requirement.  Consequently, a same-photon-assisted, passive two-photon teleportation node that claims deterministic high-dimensional teleportation through complete Bell-label readout must have $r_\Phi\ge d$ .
If \(r_\Phi<d\), the BM subroutine cannot be deterministic in this resource class.  This statement does not exclude other teleportation architectures: linear-optical qudit teleportation with additional entangled photons \cite{Goyal2014,ZhangTeleport2019} or nonlinear high-dimensional teleportation~\cite{Bianchi2025_PRA} uses different resources and is outside the theorem.

\subsection{Entanglement swapping and repeater nodes}

Entanglement swapping uses a BM on two middle systems to convert two independent entangled links into one longer entangled link~\cite{Zukowski1993,Pan1998}.  For qudits, consider ideal elementary links
\[
\ket{\Phi_d}_{A M_1}\ket{\Phi_d}_{M_2 B}.
\]
A complete BM on \(M_1M_2\) projects \(AB\) onto a maximally entangled state; the measurement outcome gives the Bell label and therefore the corresponding generalized Pauli correction.  Therefore, a full-label BM is the local operation that turns two short links into one longer link with known Pauli frame.

Let a repeater chain contain \(N_{\rm swap}\) independent swapping stations, and condition on all elementary links having been created and all photons arriving at their corresponding swapping stations.  If each station uses a Bell-label measurement with success probability \(p_{\rm BM}\), then the probability that all swaps provide complete Bell-label information is
\begin{equation}
p_{\rm chain|links}=p_{\rm BM}^{N_{\rm swap}} .
\label{eq:supp_chain_success}
\end{equation}
This simple expression is not a full repeater-rate formula with memories, multiplexing, scheduling, or loss; it is the BM subroutine factor common to swapping-based repeater protocols~\cite{Briegel1998,Kimble2008,Wehner2018}.  The theorem implies that a same-photon-assisted high-dimensional swapping station can make this factor equal to unity only if the total auxiliary Schmidt rank satisfies \(r_\Phi\ge d\).  If \(r_\Phi<d\), a station in this resource class must be probabilistic, grouped, or outside the static passive two-photon model.  This is the sense in which the auxiliary Schmidt rank becomes a node-level resource for high-dimensional photonic repeaters and network
architectures \cite{Bacco2021,Baghdasar2025}.

\subsection{Dense coding}
In \(d\)-dimensional dense coding, a sender applies one of the \(d^2\) Weyl operators \(X^pZ^q\), \(p,q \in \mathbb{Z}_d\), to one half of a maximally entangled pair, thereby encoding one of \(d^2\) messages~\cite{Bennett1992,Mattle1996}.
Recovering the full message requires complete discrimination of the
Bell label.  Therefore, in the same-photon-assisted passive model,
deterministic dense-code readout of all \(d^2\) messages requires $r_\Phi\ge d$.
A grouped analyzer may still transmit partial information, but then the relevant figure of merit is the grouped mutual information of Sec.~\ref{sec:supp_scope}, not the full dense-coding value \(2\log_2 d\) bits per entangled pair.  If failures are heralded and discarded, an ideal complete BM success probability \(p_{\rm BM}\) gives a full-message throughput factor
\begin{equation}
R_{\rm dense}=p_{\rm BM}\,2\log_2 d
\end{equation}
bits per attempted shared pair, before including source and channel
losses.

\subsection{Fusion and encoded measurements}

Fusion measurements in photonic quantum computing and network
generation are entangling measurements used to merge photonic resource
states \cite{Browne2005,Kok2007}.  When a fusion operation is literally
implemented as complete physical Bell-label readout of two
\(d\)-dimensional photons, the present rank threshold applies to that
subroutine.  However, many useful fusion protocols have a different
success criterion: they project onto a useful subspace, perform
encoded-readout operations, or accept grouped outcomes rather than
reporting all \(d^2\) physical Bell labels
\cite{Hilaire2023,Schmidt2024Fusion,Ustun2025,Yamazaki2025,Reiss2026LogicalBM}.
Those protocols are not counterexamples to the theorem; they solve a
different measurement task.  For example, high-dimensional Type-II
fusion protocols can have success probabilities that scale differently
from complete physical Bell-label readout because their target is a
fusion projection rather than a deterministic \(d^2\)-label Bell
measurement \cite{Ustun2025,Yamazaki2025}.

\end{document}